\documentclass[aps, prd, amsmath, floats, twocolumn, floatfix, nofootinbib, superscriptaddress]{revtex4-1}
\usepackage{graphicx}
\usepackage{color}
\usepackage{latexsym}
\usepackage{epsfig}
\usepackage[colorlinks]{hyperref}
\usepackage{array}
\usepackage{enumitem}
\usepackage{textcomp}
\usepackage{footmisc}
\setlist{noitemsep,leftmargin=*}

\newcommand{\be}{\begin{eqnarray}}
\newcommand{\ee}{\end{eqnarray}}
\newcommand{\bead}{\begin{eqnarray}\begin{aligned}}
\newcommand{\eead}{\end{aligned}\end{eqnarray}}
\newcommand{\beq}{\begin{equation*}}
\newcommand{\eeq}{\end{equation*}}

\newcommand{\bitem}{\begin{itemize}}
\newcommand{\eitem}{\end{itemize}}
\newcommand{\bwide}{\begin{widetext}}
\newcommand{\ewide}{\end{widetext}}

\newcommand{\dee}{\,\textrm{d}}

\begin{document}

\title{X-ray reflection spectroscopy with Kaluza-Klein black holes}

\author{Jiachen~Zhu}
\affiliation{Center for Field Theory and Particle Physics and Department of Physics, Fudan University, 200438 Shanghai, China}

\author{Askar~B.~Abdikamalov}
\affiliation{Center for Field Theory and Particle Physics and Department of Physics, Fudan University, 200438 Shanghai, China}

\author{Dimitry~Ayzenberg}
\affiliation{Center for Field Theory and Particle Physics and Department of Physics, Fudan University, 200438 Shanghai, China}

\author{Mustapha~Azreg-A\"inou}
\affiliation{Engineering Faculty, Baskent University, 06790 Ankara, Turkey}

\author{Cosimo~Bambi}
\email[Corresponding author: ]{bambi@fudan.edu.cn}
\affiliation{Center for Field Theory and Particle Physics and Department of Physics, Fudan University, 200438 Shanghai, China}

\author{Mubasher~Jamil}
\affiliation{Institute for Theoretical Physics and Cosmology, Zhejiang University of Technology, 310023 Hangzhou, China}
\affiliation{Department of Mathematics, School of Natural Sciences (SNS), National University of Sciences and Technology (NUST), H-12, 44000 Islamabad, Pakistan}
\affiliation{Canadian Quantum Research Center, 204-3002, 32 Ave, Vernon, BC, V1T 2L7, Canada}

\author{Sourabh~Nampalliwar}
\affiliation{Theoretical Astrophysics, Eberhard-Karls Universit\"at T\"ubingen, 72076 T\"ubingen, Germany}

\author{Ashutosh~Tripathi}
\affiliation{Center for Field Theory and Particle Physics and Department of Physics, Fudan University, 200438 Shanghai, China}

\author{Menglei~Zhou}
\affiliation{Center for Field Theory and Particle Physics and Department of Physics, Fudan University, 200438 Shanghai, China}

\begin{abstract}
Kaluza-Klein theory is a popular alternative theory of gravity, with both non-rotating and rotating black hole solutions known. This allows for the possibility that the theory could be observationally tested. We present a model which calculates the reflection spectrum of a black hole accretion disk system, where the black hole is described by a rotating solution of the Kaluza-Klein theory. We also use this model to analyze X-ray data from the stella-mass black hole in GRS~1915+105 and provide constraints on the free parameters of the Kaluza-Klein black holes.
\end{abstract}

\maketitle


\section{Introduction \label{s-intro}}

General relativity, the standard theory of gravitation today, has been applied to a large range of astrophysical phenomena in our Universe. Over the years, it has undergone a plethora of tests; while largely successful in the weak-field regime~\cite{will2014}, its predictions in the strong-field regime have recently become testable in a variety of ways~\cite{Bambi2015,Bambi:2017khi,LIGOScientific:2019fpa}. During this period, various shortcomings of general relativity (GR hereafter), both from the theoretical (e.g, singularities, difficulty integrating with quantum mechanics, the hierarchy problem) as well as observational point of view (e.g., dark matter, dark energy), have been exposed. Resolutions range from conservative (additional fields, extensions with GR as a limit) to radical (modifications to GR) proposals. In either case, without testable predictions it is impossible to determine if the resolution, conservative or radical, is valid.

Black holes (BHs hereafter) promise to be the strongest probes of gravity in our Universe. Their compactness means gravitational effects, of GR or of alternatives to GR, are strongest in their vicinity. Their simplicity within GR means deviations away from GR, if imprinted on BH solutions, could be easily detected. Finally, their ubiquitousness in nature means there are several potential sources to study. Thus, testing alternative theories of gravity is particularly promising with BHs.

Gravitational waves, imaging of BH shadow, and X-ray spectroscopy are the leading techniques for probing astrophysical BHs. Among these, imaging is not expected to provide very strong constraints on alternative theories~\cite{Mizuno:2018lxz}. Gravitational wave interferometry is the most promising technique, though in some cases, it is expected to be comparable to X-ray spectroscopy~\cite{CardenasAvendano:2019zxd}.\footnote{In some other cases, e.g. violations of the Einstein Equivalence Principle like variation of fundamental constants~\cite{Bambi:2013mha} or deviations from geodesic motion~\cite{Li:2019lsm}, it will be unable to provide any constraint.} In the present work, our focus is on the X-ray spectroscopy technique. Specifically, we are interested in the reflection spectrum of BH accretion disks~\cite{Reynolds:2013qqa,Bambi2015,Bambi2018}. For Kerr BHs of GR, \textsc{relxill} is the leading model for analysis of the reflection spectrum ~\cite{Garcia2013, Dauser2014}. Some of us have been involved in generalizing this model to non-Kerr metrics~\cite{relxillnk,Abdikamalov:2019yrr}. The model, \textsc{relxill\_nk}, has been applied to X-ray observations of several astrophysical BHs to place constraints on deviations away from the Kerr solution~\cite{Nampalliwar:2019iti,Liu:2019vqh,Zhou:2019kwb,Cao2017,Tripathi2018a,Xu2018,Choudhury:2018zmf,Zhou2018a,Zhou2018b,Tripathi:2018lhx,Tripathi:2019bya,Zhang:2019zsn,Zhang:2019ldz,Tripathi:2019fms}. A public version of the model is available at~\cite{relxillnkweb,relxillnkweb2}. Another interesting phenomenon that can potentially be used to probe strong-field gravity is the gyroscope precession frequency. The gravitational field of the BH causes the rotational axis of a gyroscope to precess, and the precession frequency carries unique signatures of the BH metric. Such precession frequencies have been calculated in several non-Kerr spacetimes~\cite{Rizwan:2018rgs,Rizwan:2018lht,Haroon:2017opl,Chakraborty:2016mhx}. 

One of the most interesting alternative theory of gravity proposals is the Kaluza-Klein theory. With purely classical origins back in 1919, the theory has been interpreted in a quantum mechanical as well as a string theory framework. It is a five-dimensional theory, with a compact fifth dimension. The basic ingredients include three kinds of fields: gravity, electromagnetism and a scalar field. Tests of the theory include looking for signatures in the Large Hadron Collider~\cite{Khachatryan:2010wx}, but such tests have not been very successful yet. Among astrophysical tests, in~\cite{wesson1995} the authors have analyzed how equations of motion change in Kaluza-Klein cosmology which may affect motions of galaxies. Shadows of Kaluza-Klein BHs have also been analyzed in~\cite{Amarilla:2013sj}. Gravitational waves are not expected to provide good constraints in the near future~\cite{Cardoso:2019vof} (See also~\cite{Andriot:2017oaz}). In~\cite{AzregAinou:2019ylk}, some of us study the precession of a gyroscope in the vicinity of a Kaluza-Klein BH. We can therefore ask the question: can the predictions of Kaluza-Klein theory be tested using X-ray spectroscopy? This paper presents our efforts to answer this question. 

BHs in Kaluza-Klein theory have been derived in various limits~\cite{Horowitz:2011cq}. Non-rotating spherically symmetric BHs were derived in~\cite{Dobiasch:1981vh,Chodos:1980df,Gibbons:1985ac}. Larsen (among others) found rotating BH solutions in five and four dimensions~\cite{Larsen:1999pp,Rasheed:1995zv,Matos:1996km}. BHs with squashed horizon were calculated in~\cite{Ishihara:2005dp,Wang:2006nw}. Six and higher dimension versions have also been found~\cite{Park:1995wk}. Since astrophysical BHs are mostly rotating, and X-ray spectroscopy is most suited for rapidly rotating BHs, we shall focus on rotating BHs. Rotating Kaluza-Klein BHs typically have four free parameters: mass, spin and the electric and magnetic charges. We have implemented this BH metric in the \textsc{relxill\_nk} framework, and used data from an X-ray binary to get constraints on the free parameters. 

The paper is organized as follows: in Sec.~\ref{s-metric}, we review the Kaluza-Klein theory and the BH metric. In Sec.~\ref{s-xrs} we review the theory of X-ray spectroscopy, the \textsc{relxill\_nk} framework, and describe the numerical method we used to implement the Kaluza-Klein BH metric in \textsc{relxill\_nk}. In Sec.~\ref{s-analysis}, the new model is applied to X-ray observations of an X-ray binary. Conclusions follow in Sec.~\ref{s-conclude}.

\section{The metric \label{s-metric}}


We will follow the notation developed in~\cite{AzregAinou:2019ylk} for the metric. The simplest Kaluza-Klein theory involves three fields: gravity, the dilaton and the gauge field. The action in the Einstein frame is~\cite{Overduin:1998pn}:
\be\label{eq-lag}
S = \int \sqrt{-g} \left ( \frac{R}{\kappa^2} + \frac{1}{4} e^{\sqrt{3}\kappa \sigma}F_{\alpha \beta}F^{\alpha \beta} + \frac{1}{2}\nabla^{\alpha} \sigma \nabla_{\alpha} \sigma \right )d^4x,  \nonumber
 \ee
where $\sqrt{-g}$ is the determinant of the four-dimensional metric tensor, $R$ the Ricci scalar, $F_{\alpha \beta}$ the gauge field which can be identified with the electromagnetic field, and $\sigma$ a dilaton scalar field. $\kappa$ is a constant and is equal to $\sqrt{16\pi G}$.

Although we will not derive the BH solution here, it is interesting to point out some features of the solution generating techniques. Standard methods of solving the field equations can be used to derive the non-rotating solutions. Rotating solutions on the other hand have been obtained in the following ways: for slow rotation,~\cite{Allahverdizadeh:2009ay} solve the complete field equations perturbatively, following~\cite{Horne:1992zy}; others~\cite{Horowitz:2011cq,Larsen:1999pp} boost the Kerr metric along a line to get a five-dimensional rotating BH solution.
The metric looks like this:
\bwide
\bead\label{solution}
	{\rm d}s^2=\frac{H_2}{H_1}({\rm d}\psi +A)^2-\frac{H_3}{H_2} ({\rm d}t+B)^2+H_1 \Big(  \frac{{\rm d}r^2}{\Delta}+{\rm d}\theta^2+\frac{\Delta}{H_3}\sin^2\theta \dee\phi^2 \Big),
\eead
where the one-forms are given by
\bead
	A=& - \frac{1}{H_2}\Big[  2Q(r+\frac{p-2m}{2})+\sqrt{\frac{q^3(p^2-4m^2)}{4m^2 (p+q)}}a\cos\theta  \Big]\dee t
	-\frac{1}{H_2}\Big[  2p(H_2+a^2\sin^2\theta)\cos\theta\\
	&+\sqrt{\frac{p(q^2-4m^2)}{4m^2 (p+q)^3}}[(p+q)(pr-m(p-2m))+q(p^2-4m^2)]a\sin^2\theta  \Big]\dee\phi,\\
	\label{B3}B=&\frac{(pq+4m^2)r-m(p-2m)(q-2m)}{2m(p+q)H_3}\sqrt{pq}a\sin^2\theta \dee\phi,
\eead
and
\bead\label{eq-5dmetricfuncs}
	H_1 &= r^2 +a^2\cos^2\theta +r(p-2m)+\frac{p(p-2m)(q-2m)}{2(p+q)}-\frac{p}{2m(p+q)}\sqrt{(q^2-4m^2)(p^2-4m^2)}~a\cos\theta,\\
	H_2 &= r^2 +a^2\cos^2\theta +r(q-2m)+\frac{q(p-2m)(q-2m)}{2(p+q)}+\frac{q}{2m(p+q)}\sqrt{(q^2-4m^2)(p^2-4m^2)}~a\cos\theta,\\
	H_3 &= r^2+a^2\cos^2\theta-2mr,\\ 
	\Delta &= r^2+a^2-2mr.
\eead
The solution admits four free parameters, viz. $m,a,p,q$, which are related to the physical mass $M$, the angular momentum $J$ and the electric ($Q$) and magnetic charge ($P$) respectively. The relations are given as:
\bead\label{eq-rels}
M &= \frac{p+q}{4}, \\
J &= \frac{\sqrt{pq}(pq+4m^2)}{4m(p+q)}a, \\
Q^2 &= \frac{q(q^2-4m^2)}{4(p+q)}, \\
P^2 &= \frac{p(p^2-4m^2)}{4(p+q)}.
\eead
The fifth dimension can be compactified and this results in a four-dimensional BH metric \cite{AzregAinou:2019ylk}:
\be\label{eq-4dmetric}
ds^2 = -\frac{H_3}{\rho^2}\dee t^2 -2\frac{H_4}{\rho^2}\dee t \dee \phi + \frac{\rho^2}{\Delta}\dee r^2 + \rho^2\dee \theta^2 + \left ( \frac{-H_4^2 + \rho^4 \Delta \sin^2{\theta}}{\rho^2H_3} \right )\dee \phi^2
 \ee
where $\rho = \sqrt{H_1 H_2}$ and
\bead\label{eq-4dmetricfuncs}
\frac{H_1}{M^2} &= \frac{8(b-2)(c-2)b}{(b+c)^3} + \frac{4(b-2)x}{b+c} + x^2 - \frac{2b\sqrt{(b^2-4)(c^2-4)}\alpha \cos{\theta}}{(b+c)^2} + \alpha^2\cos^2{\theta},\\
\frac{H_2}{M^2} &= \frac{8(b-2)(c-2)c}{(b+c)^3} + \frac{4(c-2)x}{b+c} + x^2 + \frac{2c\sqrt{(b^2-4)(c^2-4)}\alpha \cos{\theta}}{(b+c)^2} + \alpha^2\cos^2{\theta},\\
\frac{H_3}{M^2} &= x^2 + \alpha^2\cos^2{\theta} - \frac{8x}{b+c}, \\
\frac{H_4}{M^3} &= \frac{2\sqrt{bc}\left [ (bc+4)(b+c)x - 4(b-2)(c-2) \right ]\alpha \sin^2{\theta}}{(b+c)^3}, \\
\frac{\Delta}{M^2} &= x^2 + \alpha^2 - \frac{8x}{b+c}.
\eead
\ewide 
Here we have used dimensionless version of the free parameters, defined as $\alpha \equiv a/M$, $b \equiv p/m$, $c \equiv q/m$, and $x \equiv r/M$. Moreover, we can relate the free parameter $m$ and the physical mass $M$ using Eq.~\ref{eq-rels} and obtain 
\be\label{eq-mM}
m = 4M/(b+c).
\ee
Note that the spin parameter $\alpha$ is not always the same as the dimensionless spin parameter of the Kerr metric. Only when the electric and magnetic charges are zero, and the Kaluza-Klein metric reduces to the Kerr metric, does $\alpha$ equal the $a_*$ parameter of the Kerr solution.  

We now discuss some properties of this solution. It admits two horizons, viz.
\be
r_{\pm} = m \pm \sqrt{m^2-a^2},
\ee 
or, in terms of the dimensionless quantities,
\be\label{eq-xhorizon}
x_{\pm} = \frac{4 \pm \sqrt{16-\alpha^2(b+c)^2}}{b+c},
\ee 
and the determinant is equal to $\rho^2\sin^2{\theta}$. The non-rotating class of solutions ($\alpha=0$) was obtained in~\cite{Gibbons:1985ac}. For $b=c$, the non-rotating solution reduces to the Reissner-Nordstr\"{o}m solution of GR. The Kerr solution is recovered when $b=c=2$. However, when magnetic charge is zero, the metric does not reduce to the Kerr-Newman solution. Here, we are interested in BHs that parametrically deviate from the Kerr BHs. Since our data analysis models allow only one variable deformation parameter\footnote{The primary reason for this limitation is that the size of the FITS file (see Sec.~\ref{s-trf}) that is loaded in the computer's RAM is 1.4 GB for one deformation parameter and 42 GB for two. The former is manageable, the latter is not.}, we cannot allow both $b$ and $c$ to be free. Therefore, we consider the following two cases\footnote{Note that the case with $b=2$ and $c$ free (BHs with non-vanishing electric charge and vanishing magnetic charge) is analogous to Case 2, since the metric in Eq.~\ref{eq-4dmetric} has similar forms for the two choices.}: 
\begin{itemize}
	\item Case 1: $b=c$.
	\item Case 2: $b$ free, $c=2$.
\end{itemize}
Case~1 describes BHs with electric and magnetic charges; since the electric charge of macroscopic astrophysical black holes is thought to be negligible, this case should be seen as a toy-model and to illustrate the capability of X-ray reflection spectroscopy to test such a scenario with observations. Case~2 describes BHs with vanishing electric charge and non-vanishing magnetic charge; such a scenario is theoretically more motivated. Magnetically charged black holes cannot be neutralized with ordinary matter and the possibility of the existence of magnetically charged black holes can be seen as a prediction of string theory, which requires to be tested against observations~\cite{Stojkovic:2004hz,Liebling:2016orx}.

During the analysis, it is important to ensure the spacetime does not have pathologies. Requiring that the metric structure is preserved everywhere outside the horizon leads to bounds on the free parameters. We will determine these bounds now. First, following the definition of $Q^2$ and $P^2$ in Eq.~\ref{eq-rels}, we have the conditions $q\geq2m,\; p\geq2m$, or $b\geq2,\; c\geq2$. Using these and Eq.~\ref{eq-xhorizon} we arrive at a bound on $\alpha$:
\be\label{eq-spinbound}
	\alpha^2 < 1, \quad {\rm or} \quad -1 < \alpha < 1.
\ee
For upper bounds on $b$, we use Eq.~\ref{eq-xhorizon} and Eq.~\ref{eq-spinbound}, and get the following inequalities for the two cases defined above:
\begin{itemize}
	\item Case 1: 
		\be\label{eq:rangecase1}
			2 \leq b < \frac{2}{|\alpha|}\, .
		\ee
	\item Case 2: 
		\be\label{eq:rangecase2}
			2 \leq b < \frac{4}{|\alpha|}-2\, .
		\ee
\end{itemize}
For the allowed range of $c$, note that in Case 1 $b$ equals $c$, so $c$ shares the same range as $b$, while in Case 2 $c$ is fixed at $2$.

\section{X-ray reflection spectroscopy}
\label{s-xrs}
\subsection{Theory}

\begin{figure}[!htb]
		\centering
		\includegraphics[width=0.48\textwidth]{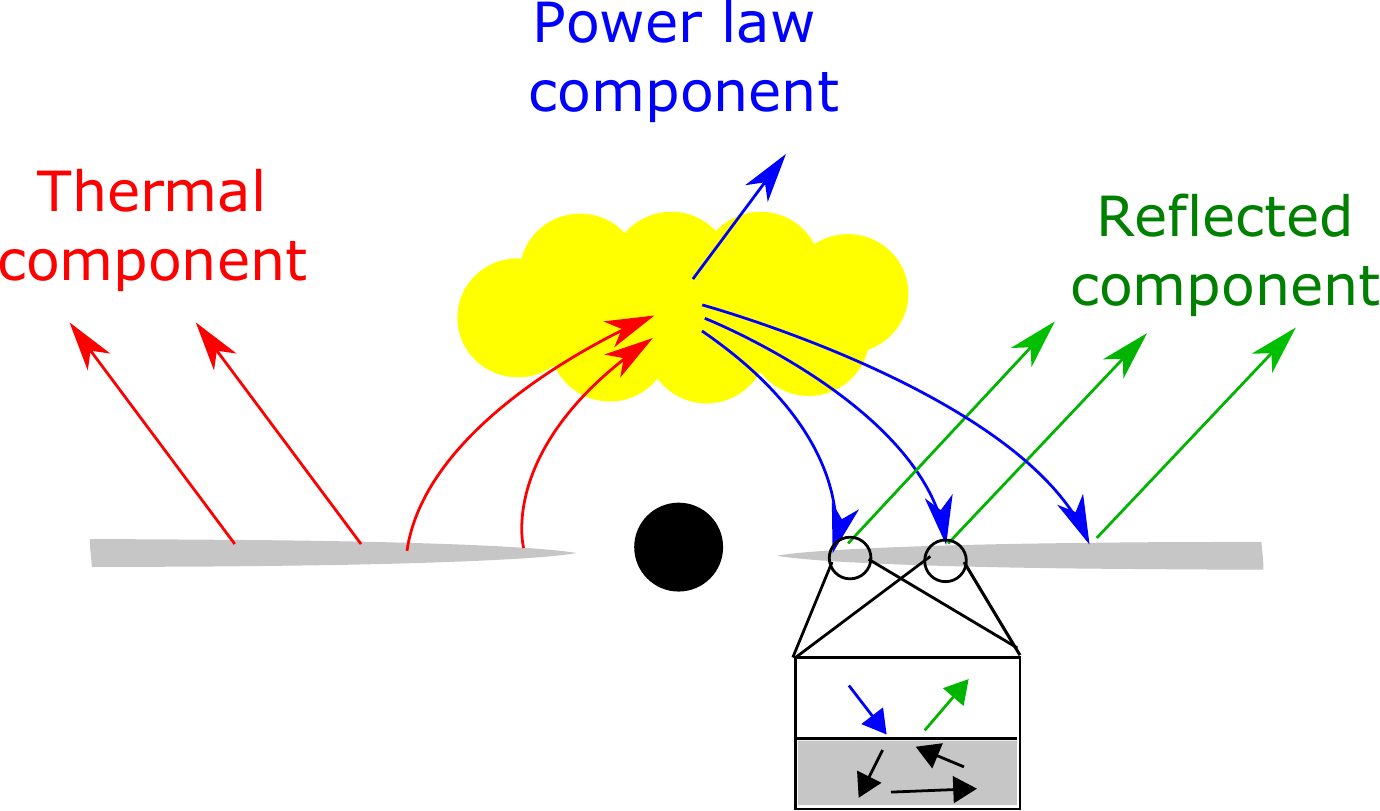}
		\caption{A cartoon of the disk-corona model. The black circle in the center indicates the BH, the disk is indicated in grey, and the corona by a yellow cloud. The structure of the corona is poorly understood so this illustration is only a guess. The arrows indicate photons and are colored according to the classification labeled on the figure and discussed in the text.}
		\label{fig:diskcorona}
\end{figure}

The standard astrophysical system we consider is a stationary BH surrounded by an accretion disk. The disk could be generated by a companion (as in stellar-mass X-ray binaries) or galactic material (as around active galactic nuclei). The typical model for a BH-accretion disk system is the disk-corona model. Fig.~\ref{fig:diskcorona} shows a cartoon of this model. The BH is assumed to be surrounded by an optically thick and geometrically thin disk in the equatorial plane~\cite{Novikov1973},
with its inner edge at some radius $r_{\textrm{in}}$, bounded by the innermost stable circular orbit (ISCO hereafter), and the outer edge at a radius $r_{\textrm{out}}$. The system also includes a ``corona''. The corona is a cloud of hot plasma (effective temperature of the order of 100~keV) whose morphology is not understood very well. The radiation spectrum includes a power-law component, produced by inverse Compton scattering of photons from the disk by the corona, a thermal continuum of blackbody radiation from the particles in the disk, and a reflection component, produced when the upscattered photons return to the disk and are reflected after reprocessing inside the disk. Our focus in the present work is on the reflection component. It is a sensitive feature and is affected not just by the BH (driving the photon from the point of emission to an observer) but also by the structure and composition of the disk, as well as the corona. 
	
\subsection{The RELXILL\_NK model \label{s-trf}}
\begin{table}[ht]
 \begin{center}
  \begin{tabular}{m{4.5cm} >{\centering\arraybackslash}m{3cm}} \\
   \hline
   \textbf{Parameter} & \textbf{Default value} \\
   \hline \hline\vspace{0.15cm}
   $q_{\rm in}$ &  $3$ \\\vspace{0.15cm}
   $q_{\rm out}$ &  $3$\\\vspace{0.15cm}
   $r_{\rm br}$ [$M$] & $15$\\\vspace{0.15cm}
   spin &  $0.998$ \\\vspace{0.15cm}
   $i$ [deg] &  $30$ \\\vspace{0.15cm}
   $r_{\rm in}$ [ISCO] & $1$\\\vspace{0.15cm}
   $r_{\rm out}$ [$M$] & $400$\\\vspace{0.15cm}
   $\Gamma$ & $2$ \\\vspace{0.15cm}
   $\log\xi$ &  $3.1$ \\\vspace{0.15cm}
   $A_{\textrm{Fe}}$ &  $1$ \\\vspace{0.15cm}
   $E_{\rm cut}$ [keV] & $300$\\\vspace{0.15cm}
   $R_f$ &  $3$ \\\vspace{0.15cm}
   $\delta$-type &  $1$ \\\vspace{0.15cm}
   $\delta$-value &  $0$ \\\vspace{0.15cm}
   $N$ & $1$\\
   \hline
   \hline
  \end{tabular}
    \caption{The parameters included in the \textsc{relxill\_nk} model and their default values. The units of the parameters, where applicable, are indicated. In particular, $r_{\rm in}$ is specified in units of ISCO by default, but can also be specified in units of $M$.}
    \label{tab:default}
 \end{center}
\end{table}
Reflection models therefore include parameters from all aspects of the disk-corona model. \textsc{relxill\_nk}~\cite{relxillnk,Abdikamalov:2019yrr,relxillnkweb,relxillnkweb2} is a suite of reflection models built for XSPEC that includes a large class of BH-disk-corona models. To illustrate, we describe the eponymous model, \textsc{relxill\_nk}. 
Tab.~\ref{tab:default} lists the parameters of the base \textsc{relxill\_nk} model as well their default values. These parameters account for the different aspects of the system. The disk's emissivity profile is modeled as a simple or broken power law as follows:
\bead
	I &\propto \frac{1}{r^{q_{\rm in}}} \qquad {\rm if} \quad r < r_{\rm br}, \nonumber \\
	I &\propto \frac{1}{r^{q_{\rm out}}} \qquad {\rm if} \quad r \geq r_{\rm br}. \nonumber
\eead
The three free parameters $q_{\rm in}$, $q_{\rm out}$ and $r_{\rm br}$ form the first three parameters of the \textsc{relxill\_nk} model. The disk is assumed to be infinitesimally thin, lying in the equatorial plane and composed of particles moving in quasi-geodesic circular orbits. Thus only two disk structure parameters $r_{\rm in}$ and $r_{\rm out}$ are needed, to account for the inner and the outer radius of the disk respectively.
The elemental constitution of the disk is assumed to follow solar abundances, except iron, which is modeled with $A_{\rm Fe}$, which is the ratio of iron content in the disk and the iron content in the sun.
The ionization of the disk is accounted with $\log\xi$ (where $\xi$ is in units of erg\:cm/s), which ranges from 0 (neutral) to 4.7 (highly ionized).
The corona is modeled with a power law, whose power law index is given by $\Gamma$ and the high energy cut-off by $E_{\rm cut}$. The latter is an observational feature and is of the order of the coronal temperature. 
The $R_f$ parameter controls the relative contributions of the coronal and the reflection spectra, and is defined as the ratio of intensity emitted towards the disk and that escaping to infinity.
The spacetime is modeled using three parameters: the BH spin $\alpha$, the deformation parameter type $\delta$-type, which can be used to switch between different deformation parameters, and the value of the deformation parameter $\delta$-value. Notably, the BH mass is not a parameter since the reflection spectrum (unlike the thermal spectrum) does not explicity depend on the BH mass.
The observer's viewing angle is modeled with $i$ and finally the strength of the spectrum is accounted with the norm $N$.


\subsection{Numerical method \label{s-trf}}
The output of the \textsc{relxill\_nk} model includes the reflection spectrum at the observer. Mathematically, this is given as
\be\label{eq-Fobs1}
F_o (\nu_o) 
= \int I_o(\nu_o, X, Y) d\tilde{\Omega}\, .
\ee
Here $I_{ o}$ is the specific intensity (for instance, in units of erg~s$^{-1}$~cm$^{-2}$~str$^{-1}$~Hz$^{-1}$) as detected by an observer. $X$ and $Y$ are the Cartesian coordinates of the image of the disk in the plane of the distant observer, and $d\tilde{\Omega} = dX dY/D^2$ is the element of the solid angle subtended by the image of the disk in the observer's sky. 
$I_{ o}$ can be related to the specific intensity at the point of emission via the Liouville's theorem: $I_o = g^3 I_e$, where $g = \nu_o/\nu_e$ is the redshift factor, $\nu_o$ is the photon frequency in the observer's frame at the point of detection, and $\nu_e$ is the photon frequency in the emitter's rest frame at the point of emission. The integration element $d\tilde{\Omega}$ which is presented in terms of variables on the observer plane can also be recast using the redshift factor and the transfer function~\cite{Cunningham1975}, where the latter is defined as follows: 
\be\label{eq-trf}
f(g^*,r_e,i) = \frac{1}{\pi r_e} g 
\sqrt{g^* (1 - g^*)} \left| \frac{\partial \left(X,Y\right)}{\partial \left(g^*,r_e\right)} \right| \, .
\ee
Here, $r_e$ is the radial coordinate at the point of emission on the disk and $g^*$ is the normalized redshift factor, defined as
\be
g^* = \frac{g - g_{\rm min}}{g_{\rm max} - g_{\rm min}} \, ,
\ee
where $g_{\rm max}=g_{\rm max}(r_e,i)$ and $g_{\rm min}=g_{\rm min}(r_e,i)$ are, respectively, the maximum and the minimum values of the redshift factor $g$ at a fixed $r_e$ and for a fixed viewing angle of the observer.
The flux can now be rewritten as
\bwide
\be\label{eq-Fobs2}
F_o (\nu_o) 
= \frac{1}{D^2} \int_{r_{\rm in}}^{r_{\rm out}} \int_0^1
\pi r_{ e} \frac{ g^2}{\sqrt{g^* (1 - g^*)}} f(g^*,r_{e},i)
I_{  e}(\nu_{ e},r_{ e},\vartheta_{ e}) \, dg^* \, dr_{ e} \, ,
\ee
\ewide
where $D$ is the distance of the observer from the source and $\vartheta_{ e}$ is the photon's direction relative to the disk at the point of emission. The $r_e$-integral ranges from the inner to the outer edge of the disk, and the $g^*$-integral ranges from 0 to 1.

\begin{figure}[!htb]
		\centering
		\includegraphics[width=0.5\textwidth]{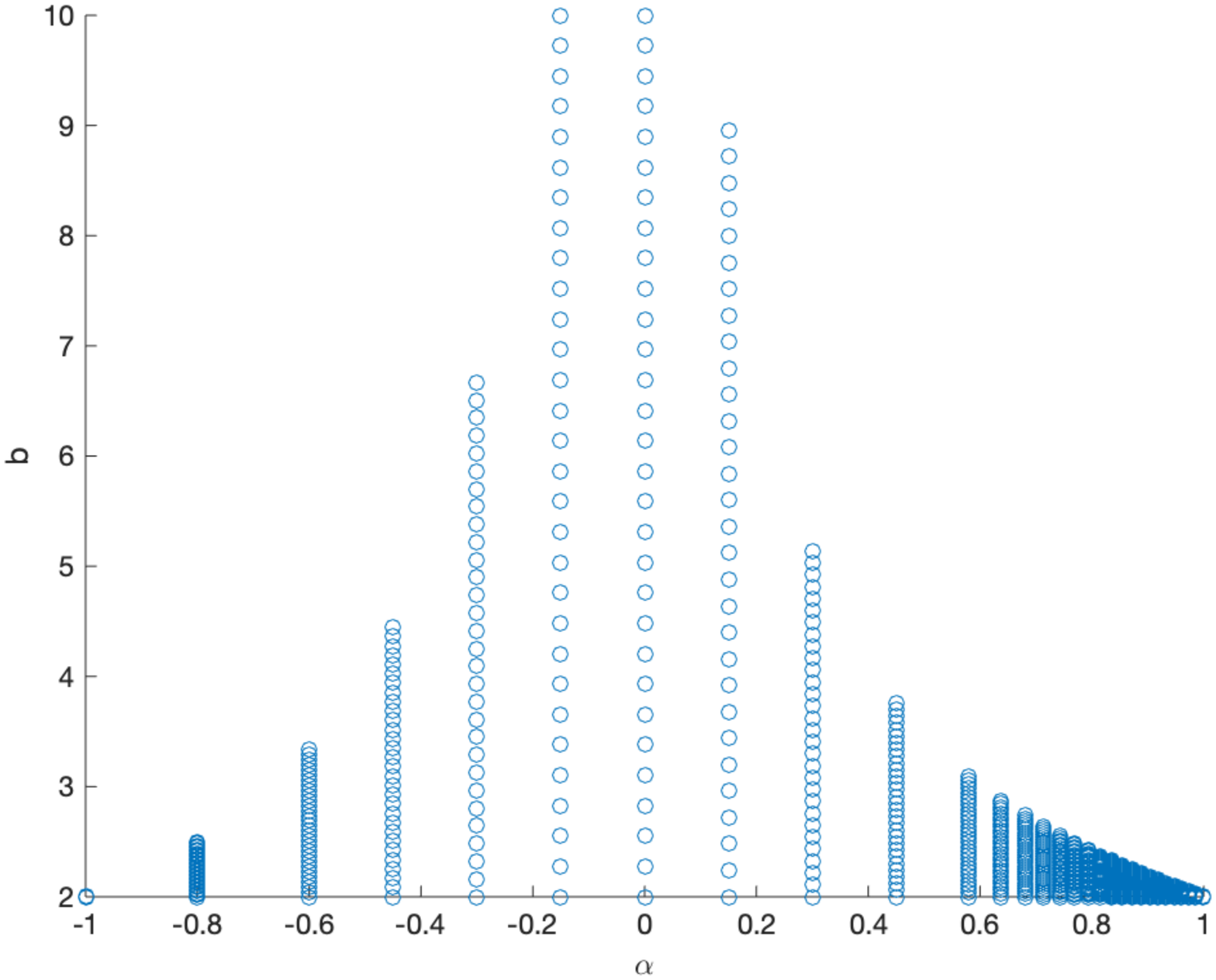}
		\caption{The grid of values, represented by blue circles, of spin $\alpha$ and deformation parameter $b$ for which the transfer functions are calculated and stored in the FITS table for Case~1. Note that the grid spacings are non-uniform in both $\alpha$ and $b$. See the text for more details.}
		\label{fig:spindefpargrid}
\end{figure}

Given the transfer function, the reflection spectrum can be readily calculated using Eq.~\ref{eq-Fobs2}. But it is computationally expensive to calculate the transfer function by tracing photons and using Eq.~\ref{eq-trf} every time the flux needs to be calculated, therefore \textsc{relxill\_nk} uses an interpolation scheme to calculate the transfer function for any $\{g^*,r_e,i\}$ using a FITS (Flexible Image Transport System) table which stores the transfer functions for some $\{g^*,r_e,i\}$. The procedure to create such a table is explained in detail in~\cite{relxillnk,Abdikamalov:2019yrr}. Here we give a brief overview. The three physical parameters spin $\alpha$, deformation parameter $b$ and the observer's viewing angle $i$, are discretized in a $30\times30\times22$ grid, respectively. Note that the grid spacing in each dimension is not necessarily uniform, e.g., the grid becomes denser as $\alpha$ increases, since the ISCO radius changes faster with increasing $\alpha$. This scheme enables sufficient resolution during interpolation while maintaining a reasonable table size.
The $b$ dimension of the grid depends on the type of deformation parameter under consideration. For the present study, our deformation parameter $b$ is bounded between $2$ and $2/|\alpha|$ in Case~1 (Eq.~\ref{eq:rangecase1}) and between $2$ and $4/|\alpha| - 2$ in Case~2 (Eq.~\ref{eq:rangecase2}). We additionally bound $b$ to be below $10$. The final $b-\alpha$ grid for Case~1 is shown in Fig.~\ref{fig:spindefpargrid} and we have a very similar grid for Case~2. Note that due to numerical complications, in some cases the bounds on $b$ are more conservative. 

At each grid point (i.e., each $\alpha$, $b$, and $i$ value in the FITS table), the accretion disk is discretized in 100 emission radii $r_e$ and 40 equally spaced $g^*$ values.\footnote{Because of the way the transfer function is defined in Eq.~\ref{eq-trf}, it goes to zero when the redshift is maximum or minimum, resulting in two branches of transfer function between $g^*=0$ and  $g^*=1$.} The emission radii grid ranges from the ISCO to $1000M$, and is non-uniform, with higher density near the ISCO. Photons are traced backwards in time from the observer plane to the disk, using the ray-tracing scheme described in~\cite{relxillnk,Abdikamalov:2019yrr}. An adaptive algorithm fine-tunes the initial location on the observer plane so that the photon, when back-traced, lands at specific $r_{\textrm{e}}$'s. For each such ``central'' photon, the code calculates the redshift, emission angle, etc. Moreover, four photons closely spaced in the observer plane are launched to calculate the Jacobian and subsequently the transfer function, using Eq.~\ref{eq-trf}. For each $r_{\textrm{e}}$, about 100 such redshifts, emission angles and transfer functions are calculated, which are then interpolated to get these quantities on the 40 equally spaced values of $g^*$, which is stored in the FITS table.

\section{Data analysis\label{s-analysis}}

In this section, we present our analysis of an X-ray observation using the \textsc{relxill\_nk} model described above.

\subsection{Review \label{ss:ana-rev}}
We chose the source GRS 1915+105 for this analysis. GRS 1915+105 (or V1487 Aquilae) is a low mass X-ray binary lying 8.6 kiloparsecs away~\cite{Reid:2014ywa}. It features one of the most massive stellar BHs known in our Galaxy. Since its last outburst in 1992, it has been a persistent source of X-rays. In previous work, we have looked at a {\em NuSTAR} and a {\em Suzaku} observation of this source. In~\cite{Zhang:2019zsn}, we used \textsc{relxill\_nk} to analyze a 2012 {\em NuSTAR} observation. This observation was difficult to fit and resulted in inconsistent values of the deformation parameter. In a follow up work~\cite{Zhang:2019ldz}, we used \textsc{relxill\_nk} to analyze a 2012 {\em Suzaku} observation, which required fewer components and resulted in a fit consistent with the Kerr metric. Note that fits to the {\em NuSTAR} observation required a thermal component, suggesting a hotter disk, unlike the {\em Suzaku} observation where no thermal component was required, suggesting a colder disk. Since the \textsc{relxill\_nk} model is based on \textsc{xillver} which assumes a cold disk, the results of the fits with the {\em Suzaku} observation can be expected to be more reliable. 

A qualitative picture emerged from previous analyses of GRS 1915+105 thus: the {\em Suzaku} observation can be fitted well with the base \textsc{relxill\_nk} model, the emissivity profile is a broken power law with very high emissivity index in the inner parts of the disk and very small in the outer parts (suggesting a ring-like corona above the accretion disk~\cite{Miniutti:2003yd,Wilkins:2011kt}), the spin is high ($\sim 0.99$), the inclination is high ($\sim 60-70$ deg) and the spacetime metric is very close to the Kerr metric. Recently, a version of \textsc{relxill\_nk} developed for thin disks of finite thickness was also used to analyze the {\em Suzaku} observation~\cite{Abdikamalov:2020oci}, which found that the finite thickness disk version of the \textsc{relxill\_nk} model provides only a marginally better fit than the base \textsc{relxill\_nk} model, which assumes infinitesimal thickness.   

\subsection{Observations and data reduction \label{s-red}}
{\em Suzaku} observed GRS 1915+105 for 117~kiloseconds on May 7, 2007 (Obs ID 402071010). During this observation, two XIS units were turned off (to preserve telemetry) and a third unit was running in the timing mode, therefore we used data from XIS1 and HXD/PIN instruments only. 

The data reduction for this observation has been described in~\cite{Zhang:2019ldz,Zhang:2019zsn}. We use the same reduced data in the analysis here. 
In particular, for the XIS1 camera a net exposure time of 28.94~ks (in the $3\times 3$ editing mode) and for the HXD/PIN a net exposure time of 53.00~ks was achieved. For the analysis, we used the 2.3 keV (since after absorption, there are insufficient photons at low energies for fitting) to 10 keV (to avoid calibration issues near the Si~K edge) energy band for XIS1 data and 12.0--55.0 keV energy band for HXD/PIN data following~\cite{Blum:2009ez}. 

 \begin{figure*}[!htb]
		\centering
		\includegraphics[width=0.49\textwidth]{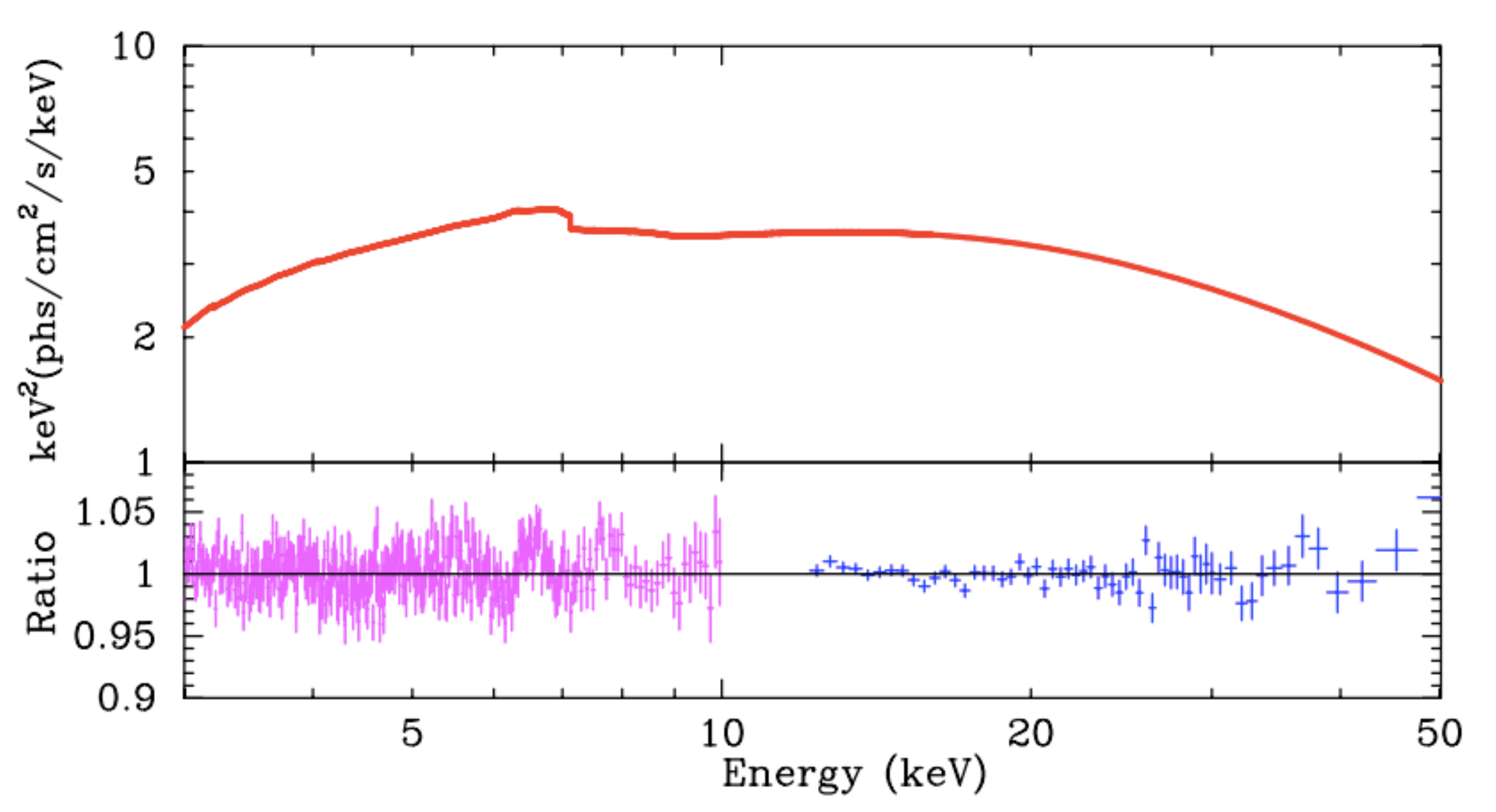}
		\includegraphics[width=0.489\textwidth]{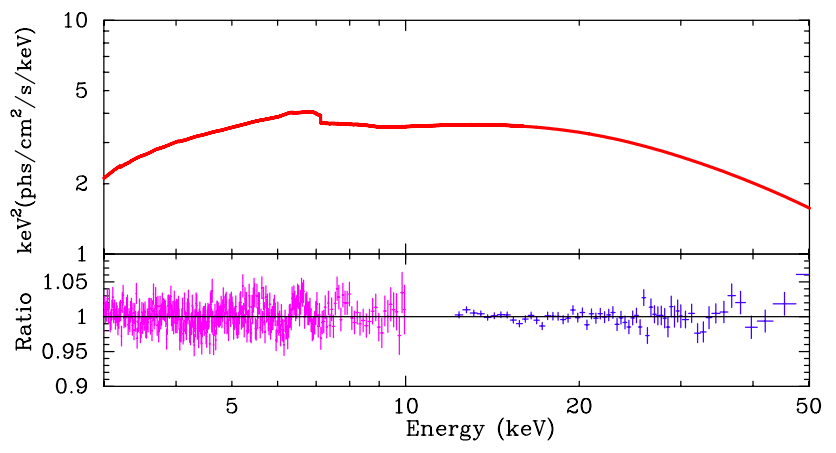}
		\caption{Best-fit models (upper quadrants) and data-to-best-fit model ratios (bottom quadrants) for Case 1 (left panel) and Case 2 (right panel). In the bottom quadrants, the XIS1 data are in magenta and the HXD/PIN data are in blue. See the text for more details.}
		\label{fig:ratio}
\end{figure*}

\subsection{Modelling and results \label{s-red}}
In our analysis, we employed XSPEC 12.10.1f. Since this observation has been analyzed before with {\sc relxill\_nk}, it was natural to guess that the best-fit model combination found previously would also work here. We thus fit the observation with the following model:

\vspace{0.07in}
{\centering
\sc {Model: tbabs*relxill\_nk},\par}
\vspace{0.05in}

\noindent where \textsc{tbabs} describes the galactic absorption~\cite{Wilms2000} and we keep the galactic column density free. The coronal and reflection spectrum are modeled with \textsc{relxill\_nk}. (The thermal spectrum does not feature in this observation, as shown in~\cite{Blum:2009ez}.) The disk emissivity profile is modeled with a broken power law. The disk inner edge lies at the ISCO, a standard assumption considering during the observation the Eddington scaled accretion luminosity was ~20\%~\cite{Blum:2009ez,Steiner:2010kd,Kulkarni:2011cy}, and the outer edge at $400M$. The results of the fit for both Case 1 and Case 2 are shown in Tab.~\ref{tab:bestfits}. The reduced $\chi^2$ is close to 1, indicating statistical agreement between model and data. The best-fit models and the data to model ratios are presented in Fig.~\ref{fig:ratio}. There are no outstanding features that appear unresolved, thus we can be confident that the model fits the data satisfactorily.
\begin{table}[ht]
 \begin{center}
  \begin{tabular}{m{3cm} >{\centering\arraybackslash}m{2.5cm} >{\centering\arraybackslash}m{2.5cm}}\\
   \hline
   \textbf{Model} & \textbf{Case 1} & \textbf{Case 2} \\
   \hline \hline
   \textsc{tbabs} \\
   $N_{\rm H}/10^{22}$~cm$^{-2}$ & $8.03^{+0.08}_{-0.05}$ & $8.05^{+0.12}_{-0.07}$  \\
   \hline
   \textsc{relxill\_nk} \\
   $q_{\rm in}$ &  $9.9^{\rm +(P)}_{-0.5}$ & $9.80^{+0.09}_{-0.82}$ \\\vspace{0.2cm}
   $q_{\rm out}$ &  $0.00^{+0.16}$ &  $0.00^{+0.14}$\\\vspace{0.2cm}
   $r_{\rm br}$ & $6.1^{+0.4}_{-0.7}$ & $6.1^{+0.7}_{-0.6}$\\\vspace{0.2cm}
   $\alpha$ &  $0.97^{+0.02}_{-0.04}$ &  $0.94^{+0.05}_{-0.02}$ \\\vspace{0.2cm}
   $i$ [deg] &  $73.56^{+1.17}_{-0.21}$ &  $73.8^{+1.0}_{-0.3}$ \\\vspace{0.2cm}
   $\Gamma$ & $2.208^{+0.051}_{-0.024}$ & $2.210^{+0.038}_{-0.020}$ \\\vspace{0.2cm}
   $\log\xi$ &  $2.77^{+0.03}_{-0.03}$ &  $2.78^{+0.06}_{-0.05}$\\\vspace{0.2cm}
   $A_{\textrm{Fe}}$ &  $0.57^{+0.06}_{-0.03}$ &  $0.56^{+0.06}_{\rm -(P)}$ \\\vspace{0.2cm}
   $E_{\rm cut}$ [keV] & $73^{+3}_{-4}$ & $74^{+4}_{-3}$\\\vspace{0.2cm}
   $R_f$ &  $0.50^{+0.03}_{-0.04}$ &  $0.50^{+0.03}_{-0.05}$ \\\vspace{0.2cm}
   $b$ &  $2.03^{+0.10}_{\rm -(P)}$ &  $2.21^{+0.06}_{\rm -(P)}$ \\
     \hline\vspace{0.2cm}
   $\chi^2/dof$  & $2304.47/2208$ & $2303.84/2208$ \\
   & $= 1.04369$ & $= 1.04341$ \\
   \hline
   \hline
  \end{tabular}
    \caption{Summary of the best-fit values from the analysis of Case~1 and Case~2. The reported uncertainty corresponds to the 90\% confidence level for one relevant parameter. $^*$ indicates that the parameter is frozen in the fit. (P) indicates that the 90\% confidence level uncertainty reaches the parameter boundary: $q_{\rm in}$ and $q_{\rm out}$ are allowed to vary in the range $[0;10]$, the lower boundary of $A_{\rm Fe}$ is 0.5, and the lower boundary of $b$ is 2.}
    \label{tab:bestfits}
 \end{center}
\end{table}
\begin{figure*}[!htb]
		\centering
		\includegraphics[width=0.49\textwidth]{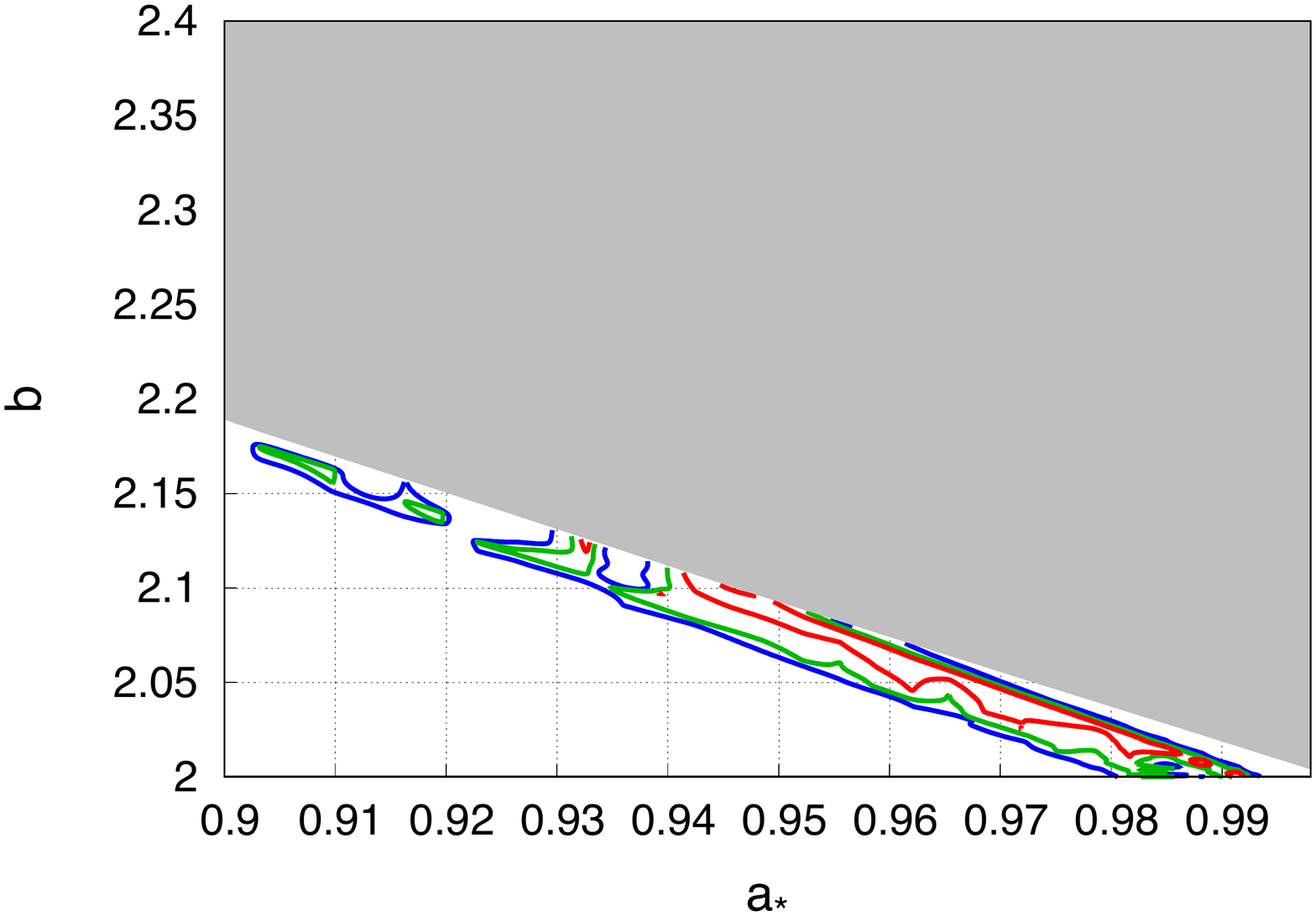}
		\includegraphics[width=0.49\textwidth]{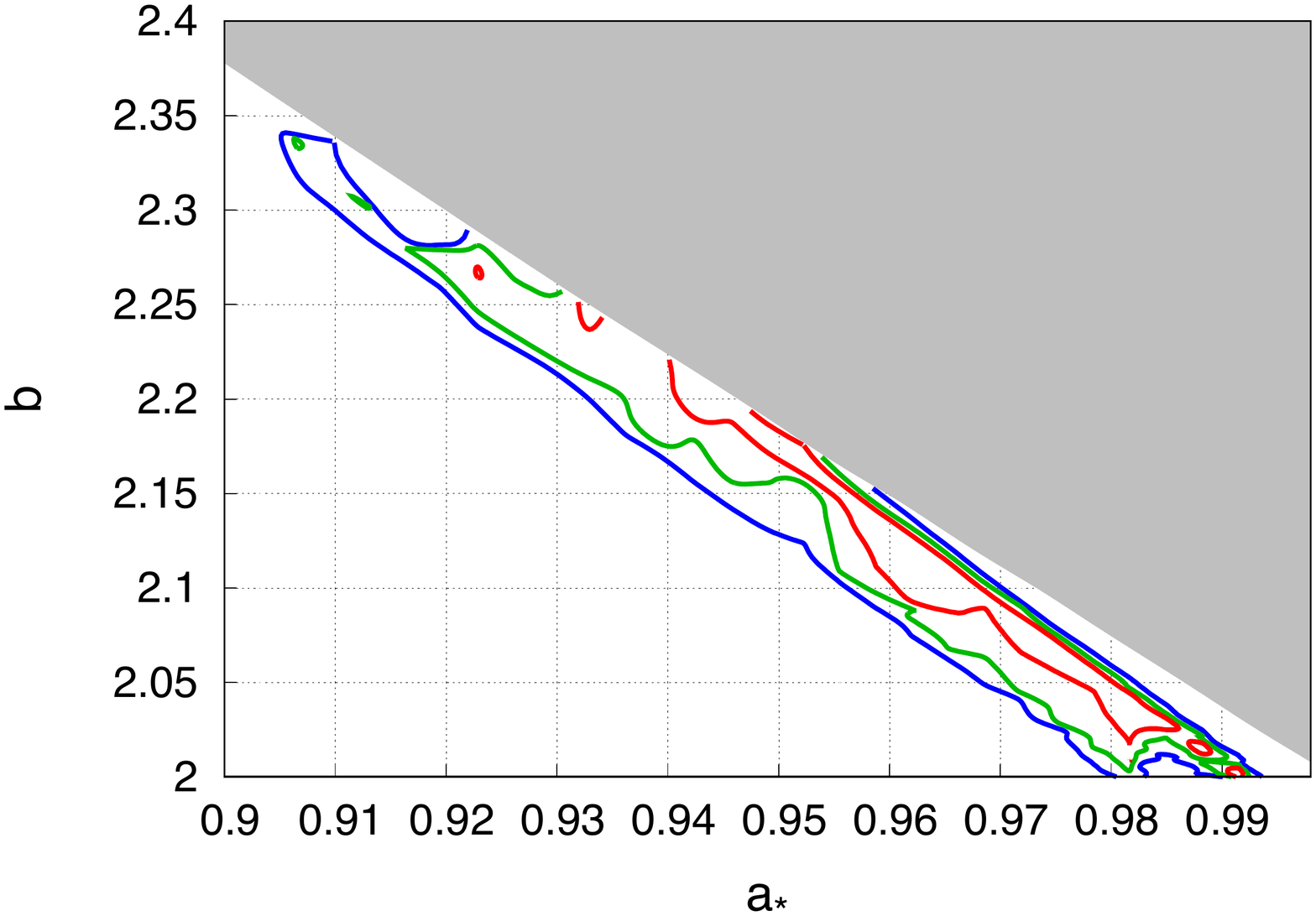}
		\vspace{-1.0cm}
		\caption{The contour plots of spin $\alpha$ vs $b$ for Case~1 (left panel) and Case~2 (right panel). The red, green and blue lines show the 68\%, 90\%, and 99\% confidence region boundaries respectively. The greyed region is pathological because outside of the parameter space described by Eq.~\ref{eq:rangecase1} and Eq.~\ref{eq:rangecase2}, respectively, and is excluded from the analysis. See the text for more details.}
		\label{fig:spindefpar}
\end{figure*}

We can compare the best-fit parameter values obtained here with their values in other analyses. The emissivity profile, for example, follows previous results with a high $q_{\rm in}$, nearly zero $q_{\rm out}$ and the break occurring near $6M$. The result can be explained within a ring-like corona above the accretion disk~\cite{Miniutti:2003yd,Wilkins:2011kt}. The spin and inclination are high, as found before. The iron abundance is below solar. 
Of course, our main interest here is the deformation parameter. We find that
\bead
	2 \leq b \leq 2.13, &\quad 2 \leq c \leq 2.13, &\quad \textrm{ Case 1}\\
	2 \leq b \leq 2.27, &\quad c \equiv 2, & \textrm{ Case 2}
\eead
within $90\%$ uncertainty for one relevant parameter. We can use Eq.~\ref{eq-rels} to translate these into constraints on the angular momentum and electric and magnetic charges of the black hole as follows:
\begin{itemize}
	\item Case 1:
	\bead
		0.87 \leq J/M^2 \leq 0.99 \, ,\\
		0 \leq Q^2/M^2 \leq 0.23 \, , \\
		0 \leq P^2/M^2 \leq 0.23 \, .
	\eead
	\item Case 2:
	\bead
		0.86 \leq J/M^2 \leq 0.99 \, ,\\
		Q^2/M^2 \equiv 0 \, , \\
		0 \leq P^2/M^2 \leq 0.34 \, .
	\eead
\end{itemize}
Since spin and deformation parameter are generally correlated, we also show contour plots of $\alpha$ vs. $b$ in Fig.~\ref{fig:spindefpar}. The red, green and blue curves are for $68\%$, $90\%$ and $99\%$ confidence, respectively, and the gray region indicates parameter space that is excluded according to Eqs.~\ref{eq:rangecase1} and~\ref{eq:rangecase2}. The correlation between $\alpha$ and $b$ is evident here. 

\section{Conclusion \label{s-conclude}}
Alternative theories of gravity have recently become observationally testable in the strong field regime. Among other techniques, X-ray reflection spectroscopy provides one of the strongest constraints on deviations from Einstein's theory. One particularly interesting example of alternatives to general relativity is the Kaluza-Klein theory. In this paper, we use one class of rotating BH solutions of Kaluza-Klein theory described in~\cite{AzregAinou:2019ylk}. We describe the astrophysical system assumed in X-ray spectroscopy analyses and an XSPEC model that calculates the reflection spectrum of this system. The XSPEC model , \textsc{relxill\_nk}, is then used to analyze a \textsl{Suzaku} observation of the stellar-mass BH system GRS 1915+105. We review the specific source, describe the specific observation, and use our model to fit the data. We find that the observation is consistent with the Kerr metric, but allows for some deviation away from it. It will be interesting to test for signatures of such deviations in other astrophysical sources in future.

{\bf Acknowledgments --}
This work was supported by the Innovation Program of the Shanghai Municipal Education Commission, Grant No.~2019-01-07-00-07-E00035, and the National Natural Science Foundation of China (NSFC), Grant No.~11973019. A.B.A. acknowledges the support from the Shanghai Government Scholarship (SGS). S.N. acknowledges support from the Excellence Initiative at Eberhard-Karls Universit\"at T\"ubingen and the Alexander von Humboldt Foundation.


\bibliography{references}

\begin{thebibliography}{65}%
\makeatletter
\providecommand \@ifxundefined [1]{%
 \@ifx{#1\undefined}
}%
\providecommand \@ifnum [1]{%
 \ifnum #1\expandafter \@firstoftwo
 \else \expandafter \@secondoftwo
 \fi
}%
\providecommand \@ifx [1]{%
 \ifx #1\expandafter \@firstoftwo
 \else \expandafter \@secondoftwo
 \fi
}%
\providecommand \natexlab [1]{#1}%
\providecommand \enquote  [1]{``#1''}%
\providecommand \bibnamefont  [1]{#1}%
\providecommand \bibfnamefont [1]{#1}%
\providecommand \citenamefont [1]{#1}%
\providecommand \href@noop [0]{\@secondoftwo}%
\providecommand \href [0]{\begingroup \@sanitize@url \@href}%
\providecommand \@href[1]{\@@startlink{#1}\@@href}%
\providecommand \@@href[1]{\endgroup#1\@@endlink}%
\providecommand \@sanitize@url [0]{\catcode `\\12\catcode `\$12\catcode
  `\&12\catcode `\#12\catcode `\^12\catcode `\_12\catcode `\%12\relax}%
\providecommand \@@startlink[1]{}%
\providecommand \@@endlink[0]{}%
\providecommand \url  [0]{\begingroup\@sanitize@url \@url }%
\providecommand \@url [1]{\endgroup\@href {#1}{\urlprefix }}%
\providecommand \urlprefix  [0]{URL }%
\providecommand \Eprint [0]{\href }%
\providecommand \doibase [0]{http://dx.doi.org/}%
\providecommand \selectlanguage [0]{\@gobble}%
\providecommand \bibinfo  [0]{\@secondoftwo}%
\providecommand \bibfield  [0]{\@secondoftwo}%
\providecommand \translation [1]{[#1]}%
\providecommand \BibitemOpen [0]{}%
\providecommand \bibitemStop [0]{}%
\providecommand \bibitemNoStop [0]{.\EOS\space}%
\providecommand \EOS [0]{\spacefactor3000\relax}%
\providecommand \BibitemShut  [1]{\csname bibitem#1\endcsname}%
\let\auto@bib@innerbib\@empty
\bibitem [{\citenamefont {Will}(2014)}]{will2014}%
  \BibitemOpen
  \bibfield  {author} {\bibinfo {author} {\bibfnamefont {C.~M.}\ \bibnamefont
  {Will}},\ }\href {\doibase 10.12942/lrr-2014-4} {\bibfield  {journal}
  {\bibinfo  {journal} {Living Rev. Rel.}\ }\textbf {\bibinfo {volume} {17}},\
  \bibinfo {pages} {4} (\bibinfo {year} {2014})},\ \Eprint
  {http://arxiv.org/abs/1403.7377} {arXiv:1403.7377 [gr-qc]} \BibitemShut
  {NoStop}%
\bibitem [{\citenamefont {Bambi}(2017{\natexlab{a}})}]{Bambi2015}%
  \BibitemOpen
  \bibfield  {author} {\bibinfo {author} {\bibfnamefont {C.}~\bibnamefont
  {Bambi}},\ }\href {\doibase 10.1103/RevModPhys.89.025001} {\bibfield
  {journal} {\bibinfo  {journal} {Rev. Mod. Phys.}\ }\textbf {\bibinfo {volume}
  {89}},\ \bibinfo {pages} {025001} (\bibinfo {year} {2017}{\natexlab{a}})},\
  \Eprint {http://arxiv.org/abs/1509.03884} {arXiv:1509.03884 [gr-qc]}
  \BibitemShut {NoStop}%
\bibitem [{\citenamefont {Bambi}(2017{\natexlab{b}})}]{Bambi:2017khi}%
  \BibitemOpen
  \bibfield  {author} {\bibinfo {author} {\bibfnamefont {C.}~\bibnamefont
  {Bambi}},\ }\href {\doibase 10.1007/978-981-10-4524-0} {\emph {\bibinfo
  {title} {{Black Holes: A Laboratory for Testing Strong Gravity}}}}\ (\bibinfo
   {publisher} {Springer},\ \bibinfo {year} {2017})\BibitemShut {NoStop}%
\bibitem [{\citenamefont {Abbott}\ \emph {et~al.}(2019)\citenamefont {Abbott}
  \emph {et~al.}}]{LIGOScientific:2019fpa}%
  \BibitemOpen
  \bibfield  {author} {\bibinfo {author} {\bibfnamefont {B.}~\bibnamefont
  {Abbott}} \emph {et~al.} (\bibinfo {collaboration} {LIGO Scientific,
  Virgo}),\ }\href {\doibase 10.1103/PhysRevD.100.104036} {\bibfield  {journal}
  {\bibinfo  {journal} {Phys. Rev. D}\ }\textbf {\bibinfo {volume} {100}},\
  \bibinfo {pages} {104036} (\bibinfo {year} {2019})},\ \Eprint
  {http://arxiv.org/abs/1903.04467} {arXiv:1903.04467 [gr-qc]} \BibitemShut
  {NoStop}%
\bibitem [{\citenamefont {Mizuno}\ \emph {et~al.}(2018)\citenamefont {Mizuno},
  \citenamefont {Younsi}, \citenamefont {Fromm}, \citenamefont {Porth},
  \citenamefont {De~Laurentis}, \citenamefont {Olivares}, \citenamefont
  {Falcke}, \citenamefont {Kramer},\ and\ \citenamefont
  {Rezzolla}}]{Mizuno:2018lxz}%
  \BibitemOpen
  \bibfield  {author} {\bibinfo {author} {\bibfnamefont {Y.}~\bibnamefont
  {Mizuno}}, \bibinfo {author} {\bibfnamefont {Z.}~\bibnamefont {Younsi}},
  \bibinfo {author} {\bibfnamefont {C.~M.}\ \bibnamefont {Fromm}}, \bibinfo
  {author} {\bibfnamefont {O.}~\bibnamefont {Porth}}, \bibinfo {author}
  {\bibfnamefont {M.}~\bibnamefont {De~Laurentis}}, \bibinfo {author}
  {\bibfnamefont {H.}~\bibnamefont {Olivares}}, \bibinfo {author}
  {\bibfnamefont {H.}~\bibnamefont {Falcke}}, \bibinfo {author} {\bibfnamefont
  {M.}~\bibnamefont {Kramer}}, \ and\ \bibinfo {author} {\bibfnamefont
  {L.}~\bibnamefont {Rezzolla}},\ }\href {\doibase 10.1038/s41550-018-0449-5}
  {\bibfield  {journal} {\bibinfo  {journal} {Nat. Astron.}\ }\textbf {\bibinfo
  {volume} {2}},\ \bibinfo {pages} {585} (\bibinfo {year} {2018})},\ \Eprint
  {http://arxiv.org/abs/1804.05812} {arXiv:1804.05812 [astro-ph.GA]}
  \BibitemShut {NoStop}%
\bibitem [{\citenamefont {Cardenas-Avendano}\ \emph {et~al.}(2019)\citenamefont
  {Cardenas-Avendano}, \citenamefont {Nampalliwar},\ and\ \citenamefont
  {Yunes}}]{CardenasAvendano:2019zxd}%
  \BibitemOpen
  \bibfield  {author} {\bibinfo {author} {\bibfnamefont {A.}~\bibnamefont
  {Cardenas-Avendano}}, \bibinfo {author} {\bibfnamefont {S.}~\bibnamefont
  {Nampalliwar}}, \ and\ \bibinfo {author} {\bibfnamefont {N.}~\bibnamefont
  {Yunes}},\ }\href@noop {} {\  (\bibinfo {year} {2019})},\ \Eprint
  {http://arxiv.org/abs/1912.08062} {arXiv:1912.08062 [gr-qc]} \BibitemShut
  {NoStop}%
\bibitem [{\citenamefont {Bambi}(2014)}]{Bambi:2013mha}%
  \BibitemOpen
  \bibfield  {author} {\bibinfo {author} {\bibfnamefont {C.}~\bibnamefont
  {Bambi}},\ }\href {\doibase 10.1088/1475-7516/2014/03/034} {\bibfield
  {journal} {\bibinfo  {journal} {JCAP}\ }\textbf {\bibinfo {volume} {03}},\
  \bibinfo {pages} {034} (\bibinfo {year} {2014})},\ \Eprint
  {http://arxiv.org/abs/1308.2470} {arXiv:1308.2470 [gr-qc]} \BibitemShut
  {NoStop}%
\bibitem [{\citenamefont {Li}\ \emph {et~al.}(2019)\citenamefont {Li},
  \citenamefont {Yan}, \citenamefont {Xue}, \citenamefont {Ren}, \citenamefont
  {Cai}, \citenamefont {Easson}, \citenamefont {Yuan},\ and\ \citenamefont
  {Zhao}}]{Li:2019lsm}%
  \BibitemOpen
  \bibfield  {author} {\bibinfo {author} {\bibfnamefont {C.}~\bibnamefont
  {Li}}, \bibinfo {author} {\bibfnamefont {S.-F.}\ \bibnamefont {Yan}},
  \bibinfo {author} {\bibfnamefont {L.}~\bibnamefont {Xue}}, \bibinfo {author}
  {\bibfnamefont {X.}~\bibnamefont {Ren}}, \bibinfo {author} {\bibfnamefont
  {Y.-F.}\ \bibnamefont {Cai}}, \bibinfo {author} {\bibfnamefont {D.~A.}\
  \bibnamefont {Easson}}, \bibinfo {author} {\bibfnamefont {Y.-F.}\
  \bibnamefont {Yuan}}, \ and\ \bibinfo {author} {\bibfnamefont
  {H.}~\bibnamefont {Zhao}},\ }\href@noop {} {\  (\bibinfo {year} {2019})},\
  \Eprint {http://arxiv.org/abs/1912.12629} {arXiv:1912.12629 [astro-ph.CO]}
  \BibitemShut {NoStop}%
\bibitem [{\citenamefont {Reynolds}(2014)}]{Reynolds:2013qqa}%
  \BibitemOpen
  \bibfield  {author} {\bibinfo {author} {\bibfnamefont {C.~S.}\ \bibnamefont
  {Reynolds}},\ }\href {\doibase 10.1007/s11214-013-0006-6} {\bibfield
  {journal} {\bibinfo  {journal} {Space Sci. Rev.}\ }\textbf {\bibinfo {volume}
  {183}},\ \bibinfo {pages} {277} (\bibinfo {year} {2014})},\ \Eprint
  {http://arxiv.org/abs/1302.3260} {arXiv:1302.3260 [astro-ph.HE]} \BibitemShut
  {NoStop}%
\bibitem [{\citenamefont {Bambi}\ \emph {et~al.}(2018)\citenamefont {Bambi},
  \citenamefont {Abdikamalov}, \citenamefont {Ayzenberg}, \citenamefont {Cao},
  \citenamefont {Liu}, \citenamefont {Nampalliwar}, \citenamefont {Tripathi},
  \citenamefont {Wang-Ji},\ and\ \citenamefont {Xu}}]{Bambi2018}%
  \BibitemOpen
  \bibfield  {author} {\bibinfo {author} {\bibfnamefont {C.}~\bibnamefont
  {Bambi}}, \bibinfo {author} {\bibfnamefont {A.~B.}\ \bibnamefont
  {Abdikamalov}}, \bibinfo {author} {\bibfnamefont {D.}~\bibnamefont
  {Ayzenberg}}, \bibinfo {author} {\bibfnamefont {Z.}~\bibnamefont {Cao}},
  \bibinfo {author} {\bibfnamefont {H.}~\bibnamefont {Liu}}, \bibinfo {author}
  {\bibfnamefont {S.}~\bibnamefont {Nampalliwar}}, \bibinfo {author}
  {\bibfnamefont {A.}~\bibnamefont {Tripathi}}, \bibinfo {author}
  {\bibfnamefont {J.}~\bibnamefont {Wang-Ji}}, \ and\ \bibinfo {author}
  {\bibfnamefont {Y.}~\bibnamefont {Xu}},\ }\href {\doibase
  10.3390/universe4070079} {\bibfield  {journal} {\bibinfo  {journal}
  {Universe}\ }\textbf {\bibinfo {volume} {4}},\ \bibinfo {pages} {79}
  (\bibinfo {year} {2018})},\ \Eprint {http://arxiv.org/abs/1806.02141}
  {arXiv:1806.02141 [gr-qc]} \BibitemShut {NoStop}%
\bibitem [{\citenamefont {Garc\'{i}a}\ \emph {et~al.}(2014)\citenamefont
  {Garc\'{i}a} \emph {et~al.}}]{Garcia2013}%
  \BibitemOpen
  \bibfield  {author} {\bibinfo {author} {\bibfnamefont {J.}~\bibnamefont
  {Garc\'{i}a}} \emph {et~al.},\ }\href {\doibase 10.1088/0004-637X/782/2/76}
  {\bibfield  {journal} {\bibinfo  {journal} {Astrophys. J.}\ }\textbf
  {\bibinfo {volume} {782}},\ \bibinfo {pages} {76} (\bibinfo {year} {2014})},\
  \Eprint {http://arxiv.org/abs/1312.3231} {arXiv:1312.3231 [astro-ph.HE]}
  \BibitemShut {NoStop}%
\bibitem [{\citenamefont {Dauser}\ \emph {et~al.}(2014)\citenamefont {Dauser}
  \emph {et~al.}}]{Dauser2014}%
  \BibitemOpen
  \bibfield  {author} {\bibinfo {author} {\bibfnamefont {T.}~\bibnamefont
  {Dauser}} \emph {et~al.},\ }\href {\doibase 10.1093/mnrasl/slu125} {\bibfield
   {journal} {\bibinfo  {journal} {Mon. Not. Roy. Astron. Soc.}\ }\textbf
  {\bibinfo {volume} {444}},\ \bibinfo {pages} {100} (\bibinfo {year}
  {2014})},\ \Eprint {http://arxiv.org/abs/1408.2347} {arXiv:1408.2347
  [astro-ph.HE]} \BibitemShut {NoStop}%
\bibitem [{\citenamefont {Bambi}\ \emph {et~al.}(2017)\citenamefont {Bambi}
  \emph {et~al.}}]{relxillnk}%
  \BibitemOpen
  \bibfield  {author} {\bibinfo {author} {\bibfnamefont {C.}~\bibnamefont
  {Bambi}} \emph {et~al.},\ }\href {\doibase 10.3847/1538-4357/aa74c0}
  {\bibfield  {journal} {\bibinfo  {journal} {Astrophys. J.}\ }\textbf
  {\bibinfo {volume} {842}},\ \bibinfo {pages} {76} (\bibinfo {year} {2017})},\
  \Eprint {http://arxiv.org/abs/1607.00596} {arXiv:1607.00596 [gr-qc]}
  \BibitemShut {NoStop}%
\bibitem [{\citenamefont {Abdikamalov}\ \emph {et~al.}(2019)\citenamefont
  {Abdikamalov}, \citenamefont {Ayzenberg}, \citenamefont {Bambi},
  \citenamefont {Dauser}, \citenamefont {Garcia},\ and\ \citenamefont
  {Nampalliwar}}]{Abdikamalov:2019yrr}%
  \BibitemOpen
  \bibfield  {author} {\bibinfo {author} {\bibfnamefont {A.~B.}\ \bibnamefont
  {Abdikamalov}}, \bibinfo {author} {\bibfnamefont {D.}~\bibnamefont
  {Ayzenberg}}, \bibinfo {author} {\bibfnamefont {C.}~\bibnamefont {Bambi}},
  \bibinfo {author} {\bibfnamefont {T.}~\bibnamefont {Dauser}}, \bibinfo
  {author} {\bibfnamefont {J.~A.}\ \bibnamefont {Garcia}}, \ and\ \bibinfo
  {author} {\bibfnamefont {S.}~\bibnamefont {Nampalliwar}},\ }\href {\doibase
  10.3847/1538-4357/ab1f89} {\bibfield  {journal} {\bibinfo  {journal}
  {Astrophys. J.}\ }\textbf {\bibinfo {volume} {878}},\ \bibinfo {pages} {91}
  (\bibinfo {year} {2019})},\ \Eprint {http://arxiv.org/abs/1902.09665}
  {arXiv:1902.09665 [gr-qc]} \BibitemShut {NoStop}%
\bibitem [{\citenamefont {Nampalliwar}\ \emph {et~al.}(2019)\citenamefont
  {Nampalliwar}, \citenamefont {Xin}, \citenamefont {Srivastava}, \citenamefont
  {Abdikamalov}, \citenamefont {Ayzenberg}, \citenamefont {Bambi},
  \citenamefont {Dauser}, \citenamefont {Garcia},\ and\ \citenamefont
  {Tripathi}}]{Nampalliwar:2019iti}%
  \BibitemOpen
  \bibfield  {author} {\bibinfo {author} {\bibfnamefont {S.}~\bibnamefont
  {Nampalliwar}}, \bibinfo {author} {\bibfnamefont {S.}~\bibnamefont {Xin}},
  \bibinfo {author} {\bibfnamefont {S.}~\bibnamefont {Srivastava}}, \bibinfo
  {author} {\bibfnamefont {A.~B.}\ \bibnamefont {Abdikamalov}}, \bibinfo
  {author} {\bibfnamefont {D.}~\bibnamefont {Ayzenberg}}, \bibinfo {author}
  {\bibfnamefont {C.}~\bibnamefont {Bambi}}, \bibinfo {author} {\bibfnamefont
  {T.}~\bibnamefont {Dauser}}, \bibinfo {author} {\bibfnamefont {J.~A.}\
  \bibnamefont {Garcia}}, \ and\ \bibinfo {author} {\bibfnamefont
  {A.}~\bibnamefont {Tripathi}},\ }\href@noop {} {\  (\bibinfo {year}
  {2019})},\ \Eprint {http://arxiv.org/abs/1903.12119} {arXiv:1903.12119
  [gr-qc]} \BibitemShut {NoStop}%
\bibitem [{\citenamefont {Liu}\ \emph {et~al.}(2019)\citenamefont {Liu},
  \citenamefont {Abdikamalov}, \citenamefont {Ayzenberg}, \citenamefont
  {Bambi}, \citenamefont {Dauser}, \citenamefont {Garcia},\ and\ \citenamefont
  {Nampalliwar}}]{Liu:2019vqh}%
  \BibitemOpen
  \bibfield  {author} {\bibinfo {author} {\bibfnamefont {H.}~\bibnamefont
  {Liu}}, \bibinfo {author} {\bibfnamefont {A.~B.}\ \bibnamefont
  {Abdikamalov}}, \bibinfo {author} {\bibfnamefont {D.}~\bibnamefont
  {Ayzenberg}}, \bibinfo {author} {\bibfnamefont {C.}~\bibnamefont {Bambi}},
  \bibinfo {author} {\bibfnamefont {T.}~\bibnamefont {Dauser}}, \bibinfo
  {author} {\bibfnamefont {J.~A.}\ \bibnamefont {Garcia}}, \ and\ \bibinfo
  {author} {\bibfnamefont {S.}~\bibnamefont {Nampalliwar}},\ }\href {\doibase
  10.1103/PhysRevD.99.123007} {\bibfield  {journal} {\bibinfo  {journal} {Phys.
  Rev. D}\ }\textbf {\bibinfo {volume} {99}},\ \bibinfo {pages} {123007}
  (\bibinfo {year} {2019})},\ \Eprint {http://arxiv.org/abs/1904.08027}
  {arXiv:1904.08027 [gr-qc]} \BibitemShut {NoStop}%
\bibitem [{\citenamefont {Zhou}\ \emph {et~al.}(2020)\citenamefont {Zhou},
  \citenamefont {Tripathi}, \citenamefont {Abdikamalov}, \citenamefont
  {Ayzenberg}, \citenamefont {Bambi}, \citenamefont {Nampalliwar},\ and\
  \citenamefont {Zhou}}]{Zhou:2019kwb}%
  \BibitemOpen
  \bibfield  {author} {\bibinfo {author} {\bibfnamefont {B.}~\bibnamefont
  {Zhou}}, \bibinfo {author} {\bibfnamefont {A.}~\bibnamefont {Tripathi}},
  \bibinfo {author} {\bibfnamefont {A.~B.}\ \bibnamefont {Abdikamalov}},
  \bibinfo {author} {\bibfnamefont {D.}~\bibnamefont {Ayzenberg}}, \bibinfo
  {author} {\bibfnamefont {C.}~\bibnamefont {Bambi}}, \bibinfo {author}
  {\bibfnamefont {S.}~\bibnamefont {Nampalliwar}}, \ and\ \bibinfo {author}
  {\bibfnamefont {M.}~\bibnamefont {Zhou}},\ }\href {\doibase
  10.1140/epjc/s10052-020-7998-3} {\bibfield  {journal} {\bibinfo  {journal}
  {Eur. Phys. J. C}\ }\textbf {\bibinfo {volume} {80}},\ \bibinfo {pages} {400}
  (\bibinfo {year} {2020})},\ \Eprint {http://arxiv.org/abs/1908.05177}
  {arXiv:1908.05177 [gr-qc]} \BibitemShut {NoStop}%
\bibitem [{\citenamefont {Cao}\ \emph {et~al.}(2018)\citenamefont {Cao},
  \citenamefont {Nampalliwar}, \citenamefont {Bambi}, \citenamefont {Dauser},\
  and\ \citenamefont {Garcia}}]{Cao2017}%
  \BibitemOpen
  \bibfield  {author} {\bibinfo {author} {\bibfnamefont {Z.}~\bibnamefont
  {Cao}}, \bibinfo {author} {\bibfnamefont {S.}~\bibnamefont {Nampalliwar}},
  \bibinfo {author} {\bibfnamefont {C.}~\bibnamefont {Bambi}}, \bibinfo
  {author} {\bibfnamefont {T.}~\bibnamefont {Dauser}}, \ and\ \bibinfo {author}
  {\bibfnamefont {J.~A.}\ \bibnamefont {Garcia}},\ }\href {\doibase
  10.1103/PhysRevLett.120.051101} {\bibfield  {journal} {\bibinfo  {journal}
  {Phys. Rev. Lett.}\ }\textbf {\bibinfo {volume} {120}},\ \bibinfo {pages}
  {051101} (\bibinfo {year} {2018})},\ \Eprint
  {http://arxiv.org/abs/1709.00219} {arXiv:1709.00219 [gr-qc]} \BibitemShut
  {NoStop}%
\bibitem [{\citenamefont {Tripathi}\ \emph {et~al.}(2018)\citenamefont
  {Tripathi}, \citenamefont {Nampalliwar}, \citenamefont {Abdikamalov},
  \citenamefont {Ayzenberg}, \citenamefont {Jiang},\ and\ \citenamefont
  {Bambi}}]{Tripathi2018a}%
  \BibitemOpen
  \bibfield  {author} {\bibinfo {author} {\bibfnamefont {A.}~\bibnamefont
  {Tripathi}}, \bibinfo {author} {\bibfnamefont {S.}~\bibnamefont
  {Nampalliwar}}, \bibinfo {author} {\bibfnamefont {A.~B.}\ \bibnamefont
  {Abdikamalov}}, \bibinfo {author} {\bibfnamefont {D.}~\bibnamefont
  {Ayzenberg}}, \bibinfo {author} {\bibfnamefont {J.}~\bibnamefont {Jiang}}, \
  and\ \bibinfo {author} {\bibfnamefont {C.}~\bibnamefont {Bambi}},\ }\href
  {\doibase 10.1103/PhysRevD.98.023018} {\bibfield  {journal} {\bibinfo
  {journal} {Phys. Rev.}\ }\textbf {\bibinfo {volume} {D98}},\ \bibinfo {pages}
  {023018} (\bibinfo {year} {2018})},\ \Eprint
  {http://arxiv.org/abs/1804.10380} {arXiv:1804.10380 [gr-qc]} \BibitemShut
  {NoStop}%
\bibitem [{\citenamefont {Xu}\ \emph {et~al.}(2018)\citenamefont {Xu},
  \citenamefont {Nampalliwar}, \citenamefont {Abdikamalov}, \citenamefont
  {Ayzenberg}, \citenamefont {Bambi}, \citenamefont {Dauser}, \citenamefont
  {Garcia},\ and\ \citenamefont {Jiang}}]{Xu2018}%
  \BibitemOpen
  \bibfield  {author} {\bibinfo {author} {\bibfnamefont {Y.}~\bibnamefont
  {Xu}}, \bibinfo {author} {\bibfnamefont {S.}~\bibnamefont {Nampalliwar}},
  \bibinfo {author} {\bibfnamefont {A.~B.}\ \bibnamefont {Abdikamalov}},
  \bibinfo {author} {\bibfnamefont {D.}~\bibnamefont {Ayzenberg}}, \bibinfo
  {author} {\bibfnamefont {C.}~\bibnamefont {Bambi}}, \bibinfo {author}
  {\bibfnamefont {T.}~\bibnamefont {Dauser}}, \bibinfo {author} {\bibfnamefont
  {J.~A.}\ \bibnamefont {Garcia}}, \ and\ \bibinfo {author} {\bibfnamefont
  {J.}~\bibnamefont {Jiang}},\ }\href {\doibase 10.3847/1538-4357/aadb9d}
  {\bibfield  {journal} {\bibinfo  {journal} {Astrophys. J.}\ }\textbf
  {\bibinfo {volume} {865}},\ \bibinfo {pages} {134} (\bibinfo {year}
  {2018})},\ \Eprint {http://arxiv.org/abs/1807.10243} {arXiv:1807.10243
  [gr-qc]} \BibitemShut {NoStop}%
\bibitem [{\citenamefont {Choudhury}\ \emph {et~al.}(2019)\citenamefont
  {Choudhury}, \citenamefont {Nampalliwar}, \citenamefont {Abdikamalov},
  \citenamefont {Ayzenberg}, \citenamefont {Bambi}, \citenamefont {Dauser},\
  and\ \citenamefont {Garcia}}]{Choudhury:2018zmf}%
  \BibitemOpen
  \bibfield  {author} {\bibinfo {author} {\bibfnamefont {K.}~\bibnamefont
  {Choudhury}}, \bibinfo {author} {\bibfnamefont {S.}~\bibnamefont
  {Nampalliwar}}, \bibinfo {author} {\bibfnamefont {A.~B.}\ \bibnamefont
  {Abdikamalov}}, \bibinfo {author} {\bibfnamefont {D.}~\bibnamefont
  {Ayzenberg}}, \bibinfo {author} {\bibfnamefont {C.}~\bibnamefont {Bambi}},
  \bibinfo {author} {\bibfnamefont {T.}~\bibnamefont {Dauser}}, \ and\ \bibinfo
  {author} {\bibfnamefont {J.~A.}\ \bibnamefont {Garcia}},\ }\href {\doibase
  10.3847/1538-4357/ab24d6} {\bibfield  {journal} {\bibinfo  {journal}
  {Astrophys. J.}\ }\textbf {\bibinfo {volume} {879}},\ \bibinfo {pages} {80}
  (\bibinfo {year} {2019})},\ \Eprint {http://arxiv.org/abs/1809.06669}
  {arXiv:1809.06669 [gr-qc]} \BibitemShut {NoStop}%
\bibitem [{\citenamefont {Zhou}\ \emph {et~al.}(2018)\citenamefont {Zhou},
  \citenamefont {Cao}, \citenamefont {Abdikamalov}, \citenamefont {Ayzenberg},
  \citenamefont {Bambi}, \citenamefont {Modesto},\ and\ \citenamefont
  {Nampalliwar}}]{Zhou2018a}%
  \BibitemOpen
  \bibfield  {author} {\bibinfo {author} {\bibfnamefont {M.}~\bibnamefont
  {Zhou}}, \bibinfo {author} {\bibfnamefont {Z.}~\bibnamefont {Cao}}, \bibinfo
  {author} {\bibfnamefont {A.}~\bibnamefont {Abdikamalov}}, \bibinfo {author}
  {\bibfnamefont {D.}~\bibnamefont {Ayzenberg}}, \bibinfo {author}
  {\bibfnamefont {C.}~\bibnamefont {Bambi}}, \bibinfo {author} {\bibfnamefont
  {L.}~\bibnamefont {Modesto}}, \ and\ \bibinfo {author} {\bibfnamefont
  {S.}~\bibnamefont {Nampalliwar}},\ }\href {\doibase
  10.1103/PhysRevD.98.024007} {\bibfield  {journal} {\bibinfo  {journal} {Phys.
  Rev.}\ }\textbf {\bibinfo {volume} {D98}},\ \bibinfo {pages} {024007}
  (\bibinfo {year} {2018})},\ \Eprint {http://arxiv.org/abs/1803.07849}
  {arXiv:1803.07849 [gr-qc]} \BibitemShut {NoStop}%
\bibitem [{\citenamefont {Zhou}\ \emph {et~al.}(2019)\citenamefont {Zhou},
  \citenamefont {Abdikamalov}, \citenamefont {Ayzenberg}, \citenamefont
  {Bambi}, \citenamefont {Modesto}, \citenamefont {Nampalliwar},\ and\
  \citenamefont {Xu}}]{Zhou2018b}%
  \BibitemOpen
  \bibfield  {author} {\bibinfo {author} {\bibfnamefont {M.}~\bibnamefont
  {Zhou}}, \bibinfo {author} {\bibfnamefont {A.}~\bibnamefont {Abdikamalov}},
  \bibinfo {author} {\bibfnamefont {D.}~\bibnamefont {Ayzenberg}}, \bibinfo
  {author} {\bibfnamefont {C.}~\bibnamefont {Bambi}}, \bibinfo {author}
  {\bibfnamefont {L.}~\bibnamefont {Modesto}}, \bibinfo {author} {\bibfnamefont
  {S.}~\bibnamefont {Nampalliwar}}, \ and\ \bibinfo {author} {\bibfnamefont
  {Y.}~\bibnamefont {Xu}},\ }\href {\doibase 10.1209/0295-5075/125/30002}
  {\bibfield  {journal} {\bibinfo  {journal} {EPL}\ }\textbf {\bibinfo {volume}
  {125}},\ \bibinfo {pages} {30002} (\bibinfo {year} {2019})}\BibitemShut
  {NoStop}%
\bibitem [{\citenamefont {Tripathi}\ \emph
  {et~al.}(2019{\natexlab{a}})\citenamefont {Tripathi}, \citenamefont
  {Nampalliwar}, \citenamefont {Abdikamalov}, \citenamefont {Ayzenberg},
  \citenamefont {Bambi}, \citenamefont {Dauser}, \citenamefont {Garcia},\ and\
  \citenamefont {Marinucci}}]{Tripathi:2018lhx}%
  \BibitemOpen
  \bibfield  {author} {\bibinfo {author} {\bibfnamefont {A.}~\bibnamefont
  {Tripathi}}, \bibinfo {author} {\bibfnamefont {S.}~\bibnamefont
  {Nampalliwar}}, \bibinfo {author} {\bibfnamefont {A.~B.}\ \bibnamefont
  {Abdikamalov}}, \bibinfo {author} {\bibfnamefont {D.}~\bibnamefont
  {Ayzenberg}}, \bibinfo {author} {\bibfnamefont {C.}~\bibnamefont {Bambi}},
  \bibinfo {author} {\bibfnamefont {T.}~\bibnamefont {Dauser}}, \bibinfo
  {author} {\bibfnamefont {J.~A.}\ \bibnamefont {Garcia}}, \ and\ \bibinfo
  {author} {\bibfnamefont {A.}~\bibnamefont {Marinucci}},\ }\href {\doibase
  10.3847/1538-4357/ab0e7e} {\bibfield  {journal} {\bibinfo  {journal}
  {Astrophys. J.}\ }\textbf {\bibinfo {volume} {875}},\ \bibinfo {pages} {56}
  (\bibinfo {year} {2019}{\natexlab{a}})},\ \Eprint
  {http://arxiv.org/abs/1811.08148} {arXiv:1811.08148 [gr-qc]} \BibitemShut
  {NoStop}%
\bibitem [{\citenamefont {Tripathi}\ \emph
  {et~al.}(2019{\natexlab{b}})\citenamefont {Tripathi} \emph
  {et~al.}}]{Tripathi:2019bya}%
  \BibitemOpen
  \bibfield  {author} {\bibinfo {author} {\bibfnamefont {A.}~\bibnamefont
  {Tripathi}} \emph {et~al.},\ }\href {\doibase 10.3847/1538-4357/ab0a00}
  {\bibfield  {journal} {\bibinfo  {journal} {Astrophys. J.}\ }\textbf
  {\bibinfo {volume} {874}},\ \bibinfo {pages} {135} (\bibinfo {year}
  {2019}{\natexlab{b}})},\ \Eprint {http://arxiv.org/abs/1901.03064}
  {arXiv:1901.03064 [gr-qc]} \BibitemShut {NoStop}%
\bibitem [{\citenamefont {Zhang}\ \emph
  {et~al.}(2019{\natexlab{a}})\citenamefont {Zhang}, \citenamefont
  {Abdikamalov}, \citenamefont {Ayzenberg}, \citenamefont {Bambi},
  \citenamefont {Dauser}, \citenamefont {Garcia},\ and\ \citenamefont
  {Nampalliwar}}]{Zhang:2019zsn}%
  \BibitemOpen
  \bibfield  {author} {\bibinfo {author} {\bibfnamefont {Y.}~\bibnamefont
  {Zhang}}, \bibinfo {author} {\bibfnamefont {A.~B.}\ \bibnamefont
  {Abdikamalov}}, \bibinfo {author} {\bibfnamefont {D.}~\bibnamefont
  {Ayzenberg}}, \bibinfo {author} {\bibfnamefont {C.}~\bibnamefont {Bambi}},
  \bibinfo {author} {\bibfnamefont {T.}~\bibnamefont {Dauser}}, \bibinfo
  {author} {\bibfnamefont {J.~A.}\ \bibnamefont {Garcia}}, \ and\ \bibinfo
  {author} {\bibfnamefont {S.}~\bibnamefont {Nampalliwar}},\ }\href {\doibase
  10.3847/1538-4357/ab0e79} {\bibfield  {journal} {\bibinfo  {journal}
  {Astrophys. J.}\ }\textbf {\bibinfo {volume} {875}},\ \bibinfo {pages} {41}
  (\bibinfo {year} {2019}{\natexlab{a}})},\ \Eprint
  {http://arxiv.org/abs/1901.06117} {arXiv:1901.06117 [gr-qc]} \BibitemShut
  {NoStop}%
\bibitem [{\citenamefont {Zhang}\ \emph
  {et~al.}(2019{\natexlab{b}})\citenamefont {Zhang}, \citenamefont
  {Abdikamalov}, \citenamefont {Ayzenberg}, \citenamefont {Bambi},\ and\
  \citenamefont {Nampalliwar}}]{Zhang:2019ldz}%
  \BibitemOpen
  \bibfield  {author} {\bibinfo {author} {\bibfnamefont {Y.}~\bibnamefont
  {Zhang}}, \bibinfo {author} {\bibfnamefont {A.~B.}\ \bibnamefont
  {Abdikamalov}}, \bibinfo {author} {\bibfnamefont {D.}~\bibnamefont
  {Ayzenberg}}, \bibinfo {author} {\bibfnamefont {C.}~\bibnamefont {Bambi}}, \
  and\ \bibinfo {author} {\bibfnamefont {S.}~\bibnamefont {Nampalliwar}},\
  }\href {\doibase 10.3847/1538-4357/ab4271} {\bibfield  {journal} {\bibinfo
  {journal} {Astrophys. J.}\ }\textbf {\bibinfo {volume} {884}},\ \bibinfo
  {pages} {147} (\bibinfo {year} {2019}{\natexlab{b}})},\ \Eprint
  {http://arxiv.org/abs/1907.03084} {arXiv:1907.03084 [gr-qc]} \BibitemShut
  {NoStop}%
\bibitem [{\citenamefont {Tripathi}\ \emph
  {et~al.}(2019{\natexlab{c}})\citenamefont {Tripathi}, \citenamefont
  {Abdikamalov}, \citenamefont {Ayzenberg}, \citenamefont {Bambi},\ and\
  \citenamefont {Nampalliwar}}]{Tripathi:2019fms}%
  \BibitemOpen
  \bibfield  {author} {\bibinfo {author} {\bibfnamefont {A.}~\bibnamefont
  {Tripathi}}, \bibinfo {author} {\bibfnamefont {A.~B.}\ \bibnamefont
  {Abdikamalov}}, \bibinfo {author} {\bibfnamefont {D.}~\bibnamefont
  {Ayzenberg}}, \bibinfo {author} {\bibfnamefont {C.}~\bibnamefont {Bambi}}, \
  and\ \bibinfo {author} {\bibfnamefont {S.}~\bibnamefont {Nampalliwar}},\
  }\href {\doibase 10.1103/PhysRevD.99.083001} {\bibfield  {journal} {\bibinfo
  {journal} {Phys. Rev. D}\ }\textbf {\bibinfo {volume} {99}},\ \bibinfo
  {pages} {083001} (\bibinfo {year} {2019}{\natexlab{c}})},\ \Eprint
  {http://arxiv.org/abs/1903.04071} {arXiv:1903.04071 [gr-qc]} \BibitemShut
  {NoStop}%
\bibitem [{rel({\natexlab{a}})}]{relxillnkweb}%
  \BibitemOpen
  \href@noop {} {\enquote {\bibinfo {title} {\textsc{relxill\_nk}},}\ }\bibinfo
  {howpublished}
  {\url{http://www.tat.physik.uni-tuebingen.de/~nampalliwar/relxill_nk/}}
  ({\natexlab{a}})\BibitemShut {NoStop}%
\bibitem [{rel({\natexlab{b}})}]{relxillnkweb2}%
  \BibitemOpen
  \href@noop {} {\enquote {\bibinfo {title} {\textsc{relxill\_nk}},}\ }\bibinfo
  {howpublished}
  {\url{http://www.physics.fudan.edu.cn/tps/people/bambi/Site/RELXILL_NK.html}}
  ({\natexlab{b}})\BibitemShut {NoStop}%
\bibitem [{\citenamefont {Rizwan}\ \emph {et~al.}(2019)\citenamefont {Rizwan},
  \citenamefont {Jamil},\ and\ \citenamefont {Jusufi}}]{Rizwan:2018rgs}%
  \BibitemOpen
  \bibfield  {author} {\bibinfo {author} {\bibfnamefont {M.}~\bibnamefont
  {Rizwan}}, \bibinfo {author} {\bibfnamefont {M.}~\bibnamefont {Jamil}}, \
  and\ \bibinfo {author} {\bibfnamefont {K.}~\bibnamefont {Jusufi}},\ }\href
  {\doibase 10.1103/PhysRevD.99.024050} {\bibfield  {journal} {\bibinfo
  {journal} {Phys. Rev. D}\ }\textbf {\bibinfo {volume} {99}},\ \bibinfo
  {pages} {024050} (\bibinfo {year} {2019})},\ \Eprint
  {http://arxiv.org/abs/1812.01331} {arXiv:1812.01331 [gr-qc]} \BibitemShut
  {NoStop}%
\bibitem [{\citenamefont {Rizwan}\ \emph {et~al.}(2018)\citenamefont {Rizwan},
  \citenamefont {Jamil},\ and\ \citenamefont {Wang}}]{Rizwan:2018lht}%
  \BibitemOpen
  \bibfield  {author} {\bibinfo {author} {\bibfnamefont {M.}~\bibnamefont
  {Rizwan}}, \bibinfo {author} {\bibfnamefont {M.}~\bibnamefont {Jamil}}, \
  and\ \bibinfo {author} {\bibfnamefont {A.}~\bibnamefont {Wang}},\ }\href
  {\doibase 10.1103/PhysRevD.98.024015} {\bibfield  {journal} {\bibinfo
  {journal} {Phys. Rev. D}\ }\textbf {\bibinfo {volume} {98}},\ \bibinfo
  {pages} {024015} (\bibinfo {year} {2018})},\ \bibinfo {note} {[Erratum:
  Phys.Rev.D 100, 029902 (2019)]},\ \Eprint {http://arxiv.org/abs/1802.04301}
  {arXiv:1802.04301 [gr-qc]} \BibitemShut {NoStop}%
\bibitem [{\citenamefont {Haroon}\ \emph {et~al.}(2018)\citenamefont {Haroon},
  \citenamefont {Jamil}, \citenamefont {Lin}, \citenamefont {Pavlovic},
  \citenamefont {Sossich},\ and\ \citenamefont {Wang}}]{Haroon:2017opl}%
  \BibitemOpen
  \bibfield  {author} {\bibinfo {author} {\bibfnamefont {S.}~\bibnamefont
  {Haroon}}, \bibinfo {author} {\bibfnamefont {M.}~\bibnamefont {Jamil}},
  \bibinfo {author} {\bibfnamefont {K.}~\bibnamefont {Lin}}, \bibinfo {author}
  {\bibfnamefont {P.}~\bibnamefont {Pavlovic}}, \bibinfo {author}
  {\bibfnamefont {M.}~\bibnamefont {Sossich}}, \ and\ \bibinfo {author}
  {\bibfnamefont {A.}~\bibnamefont {Wang}},\ }\href {\doibase
  10.1140/epjc/s10052-018-5986-7} {\bibfield  {journal} {\bibinfo  {journal}
  {Eur. Phys. J. C}\ }\textbf {\bibinfo {volume} {78}},\ \bibinfo {pages} {519}
  (\bibinfo {year} {2018})},\ \Eprint {http://arxiv.org/abs/1712.08762}
  {arXiv:1712.08762 [gr-qc]} \BibitemShut {NoStop}%
\bibitem [{\citenamefont {Chakraborty}\ \emph {et~al.}(2017)\citenamefont
  {Chakraborty}, \citenamefont {Patil}, \citenamefont {Kocherlakota},
  \citenamefont {Bhattacharyya}, \citenamefont {Joshi},\ and\ \citenamefont
  {Królak}}]{Chakraborty:2016mhx}%
  \BibitemOpen
  \bibfield  {author} {\bibinfo {author} {\bibfnamefont {C.}~\bibnamefont
  {Chakraborty}}, \bibinfo {author} {\bibfnamefont {M.}~\bibnamefont {Patil}},
  \bibinfo {author} {\bibfnamefont {P.}~\bibnamefont {Kocherlakota}}, \bibinfo
  {author} {\bibfnamefont {S.}~\bibnamefont {Bhattacharyya}}, \bibinfo {author}
  {\bibfnamefont {P.~S.}\ \bibnamefont {Joshi}}, \ and\ \bibinfo {author}
  {\bibfnamefont {A.}~\bibnamefont {Królak}},\ }\href {\doibase
  10.1103/PhysRevD.95.084024} {\bibfield  {journal} {\bibinfo  {journal} {Phys.
  Rev. D}\ }\textbf {\bibinfo {volume} {95}},\ \bibinfo {pages} {084024}
  (\bibinfo {year} {2017})},\ \Eprint {http://arxiv.org/abs/1611.08808}
  {arXiv:1611.08808 [gr-qc]} \BibitemShut {NoStop}%
\bibitem [{\citenamefont {Khachatryan}\ \emph {et~al.}(2011)\citenamefont
  {Khachatryan} \emph {et~al.}}]{Khachatryan:2010wx}%
  \BibitemOpen
  \bibfield  {author} {\bibinfo {author} {\bibfnamefont {V.}~\bibnamefont
  {Khachatryan}} \emph {et~al.} (\bibinfo {collaboration} {CMS}),\ }\href
  {\doibase 10.1016/j.physletb.2011.02.032} {\bibfield  {journal} {\bibinfo
  {journal} {Phys. Lett. B}\ }\textbf {\bibinfo {volume} {697}},\ \bibinfo
  {pages} {434} (\bibinfo {year} {2011})},\ \Eprint
  {http://arxiv.org/abs/1012.3375} {arXiv:1012.3375 [hep-ex]} \BibitemShut
  {NoStop}%
\bibitem [{\citenamefont {{Wesson}}\ and\ \citenamefont {{Ponce de
  Leon}}(1995)}]{wesson1995}%
  \BibitemOpen
  \bibfield  {author} {\bibinfo {author} {\bibfnamefont {P.~S.}\ \bibnamefont
  {{Wesson}}}\ and\ \bibinfo {author} {\bibfnamefont {J.}~\bibnamefont {{Ponce
  de Leon}}},\ }\href@noop {} {\bibfield  {journal} {\bibinfo  {journal}
  {Astronomy \& Astrophysics}\ }\textbf {\bibinfo {volume} {294}},\ \bibinfo
  {pages} {1} (\bibinfo {year} {1995})}\BibitemShut {NoStop}%
\bibitem [{\citenamefont {Amarilla}\ and\ \citenamefont
  {Eiroa}(2013)}]{Amarilla:2013sj}%
  \BibitemOpen
  \bibfield  {author} {\bibinfo {author} {\bibfnamefont {L.}~\bibnamefont
  {Amarilla}}\ and\ \bibinfo {author} {\bibfnamefont {E.~F.}\ \bibnamefont
  {Eiroa}},\ }\href {\doibase 10.1103/PhysRevD.87.044057} {\bibfield  {journal}
  {\bibinfo  {journal} {Phys. Rev. D}\ }\textbf {\bibinfo {volume} {87}},\
  \bibinfo {pages} {044057} (\bibinfo {year} {2013})},\ \Eprint
  {http://arxiv.org/abs/1301.0532} {arXiv:1301.0532 [gr-qc]} \BibitemShut
  {NoStop}%
\bibitem [{\citenamefont {Cardoso}\ \emph {et~al.}(2019)\citenamefont
  {Cardoso}, \citenamefont {Gualtieri},\ and\ \citenamefont
  {Moore}}]{Cardoso:2019vof}%
  \BibitemOpen
  \bibfield  {author} {\bibinfo {author} {\bibfnamefont {V.}~\bibnamefont
  {Cardoso}}, \bibinfo {author} {\bibfnamefont {L.}~\bibnamefont {Gualtieri}},
  \ and\ \bibinfo {author} {\bibfnamefont {C.~J.}\ \bibnamefont {Moore}},\
  }\href {\doibase 10.1103/PhysRevD.100.124037} {\bibfield  {journal} {\bibinfo
   {journal} {Phys. Rev. D}\ }\textbf {\bibinfo {volume} {100}},\ \bibinfo
  {pages} {124037} (\bibinfo {year} {2019})},\ \Eprint
  {http://arxiv.org/abs/1910.09557} {arXiv:1910.09557 [gr-qc]} \BibitemShut
  {NoStop}%
\bibitem [{\citenamefont {Andriot}\ and\ \citenamefont
  {Lucena~Gómez}(2017)}]{Andriot:2017oaz}%
  \BibitemOpen
  \bibfield  {author} {\bibinfo {author} {\bibfnamefont {D.}~\bibnamefont
  {Andriot}}\ and\ \bibinfo {author} {\bibfnamefont {G.}~\bibnamefont
  {Lucena~Gómez}},\ }\href {\doibase 10.1088/1475-7516/2017/06/048} {\bibfield
   {journal} {\bibinfo  {journal} {JCAP}\ }\textbf {\bibinfo {volume} {06}},\
  \bibinfo {pages} {048} (\bibinfo {year} {2017})},\ \bibinfo {note} {[Erratum:
  JCAP 05, E01 (2019)]},\ \Eprint {http://arxiv.org/abs/1704.07392}
  {arXiv:1704.07392 [hep-th]} \BibitemShut {NoStop}%
\bibitem [{\citenamefont {Azreg-Aïnou}\ \emph {et~al.}(2020)\citenamefont
  {Azreg-Aïnou}, \citenamefont {Jamil},\ and\ \citenamefont
  {Lin}}]{AzregAinou:2019ylk}%
  \BibitemOpen
  \bibfield  {author} {\bibinfo {author} {\bibfnamefont {M.}~\bibnamefont
  {Azreg-Aïnou}}, \bibinfo {author} {\bibfnamefont {M.}~\bibnamefont {Jamil}},
  \ and\ \bibinfo {author} {\bibfnamefont {K.}~\bibnamefont {Lin}},\ }\href
  {\doibase 10.1088/1674-1137/44/6/065101} {\bibfield  {journal} {\bibinfo
  {journal} {Chin. Phys. C}\ }\textbf {\bibinfo {volume} {44}},\ \bibinfo
  {pages} {065101} (\bibinfo {year} {2020})},\ \Eprint
  {http://arxiv.org/abs/1907.01394} {arXiv:1907.01394 [gr-qc]} \BibitemShut
  {NoStop}%
\bibitem [{\citenamefont {Horowitz}\ and\ \citenamefont
  {Wiseman}(2012)}]{Horowitz:2011cq}%
  \BibitemOpen
  \bibfield  {author} {\bibinfo {author} {\bibfnamefont {G.~T.}\ \bibnamefont
  {Horowitz}}\ and\ \bibinfo {author} {\bibfnamefont {T.}~\bibnamefont
  {Wiseman}},\ }\enquote {\bibinfo {title} {{General black holes in
  Kaluza--Klein theory}},}\ in\ \href@noop {} {\emph {\bibinfo {booktitle}
  {{Black holes in higher dimensions}}}}\ (\bibinfo {year} {2012})\ pp.\
  \bibinfo {pages} {69--98},\ \Eprint {http://arxiv.org/abs/1107.5563}
  {arXiv:1107.5563 [gr-qc]} \BibitemShut {NoStop}%
\bibitem [{\citenamefont {Dobiasch}\ and\ \citenamefont
  {Maison}(1982)}]{Dobiasch:1981vh}%
  \BibitemOpen
  \bibfield  {author} {\bibinfo {author} {\bibfnamefont {P.}~\bibnamefont
  {Dobiasch}}\ and\ \bibinfo {author} {\bibfnamefont {D.}~\bibnamefont
  {Maison}},\ }\href {\doibase 10.1007/BF00756059} {\bibfield  {journal}
  {\bibinfo  {journal} {Gen. Rel. Grav.}\ }\textbf {\bibinfo {volume} {14}},\
  \bibinfo {pages} {231} (\bibinfo {year} {1982})}\BibitemShut {NoStop}%
\bibitem [{\citenamefont {Chodos}\ and\ \citenamefont
  {Detweiler}(1982)}]{Chodos:1980df}%
  \BibitemOpen
  \bibfield  {author} {\bibinfo {author} {\bibfnamefont {A.}~\bibnamefont
  {Chodos}}\ and\ \bibinfo {author} {\bibfnamefont {S.~L.}\ \bibnamefont
  {Detweiler}},\ }\href {\doibase 10.1007/BF00756803} {\bibfield  {journal}
  {\bibinfo  {journal} {Gen. Rel. Grav.}\ }\textbf {\bibinfo {volume} {14}},\
  \bibinfo {pages} {879} (\bibinfo {year} {1982})}\BibitemShut {NoStop}%
\bibitem [{\citenamefont {Gibbons}\ and\ \citenamefont
  {Wiltshire}(1986)}]{Gibbons:1985ac}%
  \BibitemOpen
  \bibfield  {author} {\bibinfo {author} {\bibfnamefont {G.}~\bibnamefont
  {Gibbons}}\ and\ \bibinfo {author} {\bibfnamefont {D.}~\bibnamefont
  {Wiltshire}},\ }\href {\doibase 10.1016/S0003-4916(86)80012-4} {\bibfield
  {journal} {\bibinfo  {journal} {Annals Phys.}\ }\textbf {\bibinfo {volume}
  {167}},\ \bibinfo {pages} {201} (\bibinfo {year} {1986})},\ \bibinfo {note}
  {[Erratum: Annals Phys. 176, 393 (1987)]}\BibitemShut {NoStop}%
\bibitem [{\citenamefont {Larsen}(2000)}]{Larsen:1999pp}%
  \BibitemOpen
  \bibfield  {author} {\bibinfo {author} {\bibfnamefont {F.}~\bibnamefont
  {Larsen}},\ }\href {\doibase 10.1016/S0550-3213(00)00064-X} {\bibfield
  {journal} {\bibinfo  {journal} {Nucl. Phys. B}\ }\textbf {\bibinfo {volume}
  {575}},\ \bibinfo {pages} {211} (\bibinfo {year} {2000})},\ \Eprint
  {http://arxiv.org/abs/hep-th/9909102} {arXiv:hep-th/9909102} \BibitemShut
  {NoStop}%
\bibitem [{\citenamefont {Rasheed}(1995)}]{Rasheed:1995zv}%
  \BibitemOpen
  \bibfield  {author} {\bibinfo {author} {\bibfnamefont {D.}~\bibnamefont
  {Rasheed}},\ }\href {\doibase 10.1016/0550-3213(95)00396-A} {\bibfield
  {journal} {\bibinfo  {journal} {Nucl. Phys. B}\ }\textbf {\bibinfo {volume}
  {454}},\ \bibinfo {pages} {379} (\bibinfo {year} {1995})},\ \Eprint
  {http://arxiv.org/abs/hep-th/9505038} {arXiv:hep-th/9505038} \BibitemShut
  {NoStop}%
\bibitem [{\citenamefont {Matos}\ and\ \citenamefont
  {Mora}(1997)}]{Matos:1996km}%
  \BibitemOpen
  \bibfield  {author} {\bibinfo {author} {\bibfnamefont {T.}~\bibnamefont
  {Matos}}\ and\ \bibinfo {author} {\bibfnamefont {C.}~\bibnamefont {Mora}},\
  }\href {\doibase 10.1088/0264-9381/14/8/027} {\bibfield  {journal} {\bibinfo
  {journal} {Class. Quant. Grav.}\ }\textbf {\bibinfo {volume} {14}},\ \bibinfo
  {pages} {2331} (\bibinfo {year} {1997})},\ \Eprint
  {http://arxiv.org/abs/hep-th/9610013} {arXiv:hep-th/9610013} \BibitemShut
  {NoStop}%
\bibitem [{\citenamefont {Ishihara}\ and\ \citenamefont
  {Matsuno}(2006)}]{Ishihara:2005dp}%
  \BibitemOpen
  \bibfield  {author} {\bibinfo {author} {\bibfnamefont {H.}~\bibnamefont
  {Ishihara}}\ and\ \bibinfo {author} {\bibfnamefont {K.}~\bibnamefont
  {Matsuno}},\ }\href {\doibase 10.1143/PTP.116.417} {\bibfield  {journal}
  {\bibinfo  {journal} {Prog. Theor. Phys.}\ }\textbf {\bibinfo {volume}
  {116}},\ \bibinfo {pages} {417} (\bibinfo {year} {2006})},\ \Eprint
  {http://arxiv.org/abs/hep-th/0510094} {arXiv:hep-th/0510094} \BibitemShut
  {NoStop}%
\bibitem [{\citenamefont {Wang}(2006)}]{Wang:2006nw}%
  \BibitemOpen
  \bibfield  {author} {\bibinfo {author} {\bibfnamefont {T.}~\bibnamefont
  {Wang}},\ }\href {\doibase 10.1016/j.nuclphysb.2006.09.001} {\bibfield
  {journal} {\bibinfo  {journal} {Nucl. Phys. B}\ }\textbf {\bibinfo {volume}
  {756}},\ \bibinfo {pages} {86} (\bibinfo {year} {2006})},\ \Eprint
  {http://arxiv.org/abs/hep-th/0605048} {arXiv:hep-th/0605048} \BibitemShut
  {NoStop}%
\bibitem [{\citenamefont {Park}(1998)}]{Park:1995wk}%
  \BibitemOpen
  \bibfield  {author} {\bibinfo {author} {\bibfnamefont {J.}~\bibnamefont
  {Park}},\ }\href {\doibase 10.1088/0264-9381/15/4/006} {\bibfield  {journal}
  {\bibinfo  {journal} {Class. Quant. Grav.}\ }\textbf {\bibinfo {volume}
  {15}},\ \bibinfo {pages} {775} (\bibinfo {year} {1998})},\ \Eprint
  {http://arxiv.org/abs/hep-th/9503084} {arXiv:hep-th/9503084} \BibitemShut
  {NoStop}%
\bibitem [{\citenamefont {Overduin}\ and\ \citenamefont
  {Wesson}(1997)}]{Overduin:1998pn}%
  \BibitemOpen
  \bibfield  {author} {\bibinfo {author} {\bibfnamefont {J.}~\bibnamefont
  {Overduin}}\ and\ \bibinfo {author} {\bibfnamefont {P.}~\bibnamefont
  {Wesson}},\ }\href {\doibase 10.1016/S0370-1573(96)00046-4} {\bibfield
  {journal} {\bibinfo  {journal} {Phys. Rept.}\ }\textbf {\bibinfo {volume}
  {283}},\ \bibinfo {pages} {303} (\bibinfo {year} {1997})},\ \Eprint
  {http://arxiv.org/abs/gr-qc/9805018} {arXiv:gr-qc/9805018} \BibitemShut
  {NoStop}%
\bibitem [{\citenamefont {Allahverdizadeh}\ and\ \citenamefont
  {Matsuno}(2010)}]{Allahverdizadeh:2009ay}%
  \BibitemOpen
  \bibfield  {author} {\bibinfo {author} {\bibfnamefont {M.}~\bibnamefont
  {Allahverdizadeh}}\ and\ \bibinfo {author} {\bibfnamefont {K.}~\bibnamefont
  {Matsuno}},\ }\href {\doibase 10.1103/PhysRevD.81.044001} {\bibfield
  {journal} {\bibinfo  {journal} {Phys. Rev. D}\ }\textbf {\bibinfo {volume}
  {81}},\ \bibinfo {pages} {044001} (\bibinfo {year} {2010})},\ \Eprint
  {http://arxiv.org/abs/0908.2484} {arXiv:0908.2484 [hep-th]} \BibitemShut
  {NoStop}%
\bibitem [{\citenamefont {Horne}\ and\ \citenamefont
  {Horowitz}(1992)}]{Horne:1992zy}%
  \BibitemOpen
  \bibfield  {author} {\bibinfo {author} {\bibfnamefont {J.~H.}\ \bibnamefont
  {Horne}}\ and\ \bibinfo {author} {\bibfnamefont {G.~T.}\ \bibnamefont
  {Horowitz}},\ }\href {\doibase 10.1103/PhysRevD.46.1340} {\bibfield
  {journal} {\bibinfo  {journal} {Phys. Rev. D}\ }\textbf {\bibinfo {volume}
  {46}},\ \bibinfo {pages} {1340} (\bibinfo {year} {1992})},\ \Eprint
  {http://arxiv.org/abs/hep-th/9203083} {arXiv:hep-th/9203083} \BibitemShut
  {NoStop}%
\bibitem [{\citenamefont {Stojkovic}\ and\ \citenamefont
  {Freese}(2005)}]{Stojkovic:2004hz}%
  \BibitemOpen
  \bibfield  {author} {\bibinfo {author} {\bibfnamefont {D.}~\bibnamefont
  {Stojkovic}}\ and\ \bibinfo {author} {\bibfnamefont {K.}~\bibnamefont
  {Freese}},\ }\href {\doibase 10.1016/j.physletb.2004.12.019} {\bibfield
  {journal} {\bibinfo  {journal} {Phys. Lett. B}\ }\textbf {\bibinfo {volume}
  {606}},\ \bibinfo {pages} {251} (\bibinfo {year} {2005})},\ \Eprint
  {http://arxiv.org/abs/hep-ph/0403248} {arXiv:hep-ph/0403248} \BibitemShut
  {NoStop}%
\bibitem [{\citenamefont {Liebling}\ and\ \citenamefont
  {Palenzuela}(2016)}]{Liebling:2016orx}%
  \BibitemOpen
  \bibfield  {author} {\bibinfo {author} {\bibfnamefont {S.~L.}\ \bibnamefont
  {Liebling}}\ and\ \bibinfo {author} {\bibfnamefont {C.}~\bibnamefont
  {Palenzuela}},\ }\href {\doibase 10.1103/PhysRevD.94.064046} {\bibfield
  {journal} {\bibinfo  {journal} {Phys. Rev. D}\ }\textbf {\bibinfo {volume}
  {94}},\ \bibinfo {pages} {064046} (\bibinfo {year} {2016})},\ \Eprint
  {http://arxiv.org/abs/1607.02140} {arXiv:1607.02140 [gr-qc]} \BibitemShut
  {NoStop}%
\bibitem [{\citenamefont {Novikov}\ and\ \citenamefont
  {Thorne}(1973)}]{Novikov1973}%
  \BibitemOpen
  \bibfield  {author} {\bibinfo {author} {\bibfnamefont {I.~D.}\ \bibnamefont
  {Novikov}}\ and\ \bibinfo {author} {\bibfnamefont {K.~S.}\ \bibnamefont
  {Thorne}},\ }in\ \href@noop {} {\emph {\bibinfo {booktitle} {{Proceedings,
  Ecole d'Et\'{e} de Physique Th\'{e}orique: Les Astres Occlus: Les Houches,
  France, August, 1972}}}}\ (\bibinfo {year} {1973})\ pp.\ \bibinfo {pages}
  {343--550}\BibitemShut {NoStop}%
\bibitem [{\citenamefont {Cunningham}(1975)}]{Cunningham1975}%
  \BibitemOpen
  \bibfield  {author} {\bibinfo {author} {\bibfnamefont {C.~T.}\ \bibnamefont
  {Cunningham}},\ }\href {\doibase 10.1086/154033} {\bibfield  {journal}
  {\bibinfo  {journal} {Astrophys. J.}\ }\textbf {\bibinfo {volume} {202}},\
  \bibinfo {pages} {788} (\bibinfo {year} {1975})}\BibitemShut {NoStop}%
\bibitem [{\citenamefont {Reid}\ \emph {et~al.}(2014)\citenamefont {Reid},
  \citenamefont {McClintock}, \citenamefont {Steiner}, \citenamefont {Steeghs},
  \citenamefont {Remillard}, \citenamefont {Dhawan},\ and\ \citenamefont
  {Narayan}}]{Reid:2014ywa}%
  \BibitemOpen
  \bibfield  {author} {\bibinfo {author} {\bibfnamefont {M.}~\bibnamefont
  {Reid}}, \bibinfo {author} {\bibfnamefont {J.}~\bibnamefont {McClintock}},
  \bibinfo {author} {\bibfnamefont {J.}~\bibnamefont {Steiner}}, \bibinfo
  {author} {\bibfnamefont {D.}~\bibnamefont {Steeghs}}, \bibinfo {author}
  {\bibfnamefont {R.}~\bibnamefont {Remillard}}, \bibinfo {author}
  {\bibfnamefont {V.}~\bibnamefont {Dhawan}}, \ and\ \bibinfo {author}
  {\bibfnamefont {R.}~\bibnamefont {Narayan}},\ }\href {\doibase
  10.1088/0004-637X/796/1/2} {\bibfield  {journal} {\bibinfo  {journal}
  {Astrophys. J.}\ }\textbf {\bibinfo {volume} {796}},\ \bibinfo {pages} {2}
  (\bibinfo {year} {2014})},\ \Eprint {http://arxiv.org/abs/1409.2453}
  {arXiv:1409.2453 [astro-ph.GA]} \BibitemShut {NoStop}%
\bibitem [{\citenamefont {Miniutti}\ \emph {et~al.}(2003)\citenamefont
  {Miniutti}, \citenamefont {Fabian}, \citenamefont {Goyder},\ and\
  \citenamefont {Lasenby}}]{Miniutti:2003yd}%
  \BibitemOpen
  \bibfield  {author} {\bibinfo {author} {\bibfnamefont {G.}~\bibnamefont
  {Miniutti}}, \bibinfo {author} {\bibfnamefont {A.}~\bibnamefont {Fabian}},
  \bibinfo {author} {\bibfnamefont {R.}~\bibnamefont {Goyder}}, \ and\ \bibinfo
  {author} {\bibfnamefont {A.}~\bibnamefont {Lasenby}},\ }\href {\doibase
  10.1046/j.1365-8711.2003.06988.x} {\bibfield  {journal} {\bibinfo  {journal}
  {Mon. Not. Roy. Astron. Soc.}\ }\textbf {\bibinfo {volume} {344}},\ \bibinfo
  {pages} {L22} (\bibinfo {year} {2003})},\ \Eprint
  {http://arxiv.org/abs/astro-ph/0307163} {arXiv:astro-ph/0307163} \BibitemShut
  {NoStop}%
\bibitem [{\citenamefont {Wilkins}\ and\ \citenamefont
  {Fabian}(2011)}]{Wilkins:2011kt}%
  \BibitemOpen
  \bibfield  {author} {\bibinfo {author} {\bibfnamefont {D.}~\bibnamefont
  {Wilkins}}\ and\ \bibinfo {author} {\bibfnamefont {A.}~\bibnamefont
  {Fabian}},\ }\href {\doibase 10.1111/j.1365-2966.2011.18458.x} {\bibfield
  {journal} {\bibinfo  {journal} {Mon. Not. Roy. Astron. Soc.}\ }\textbf
  {\bibinfo {volume} {414}},\ \bibinfo {pages} {1269} (\bibinfo {year}
  {2011})},\ \Eprint {http://arxiv.org/abs/1102.0433} {arXiv:1102.0433
  [astro-ph.HE]} \BibitemShut {NoStop}%
\bibitem [{\citenamefont {Abdikamalov}\ \emph {et~al.}(2020)\citenamefont
  {Abdikamalov}, \citenamefont {Ayzenberg}, \citenamefont {Bambi},
  \citenamefont {Dauser}, \citenamefont {Garcia}, \citenamefont {Nampalliwar},
  \citenamefont {Tripathi},\ and\ \citenamefont {Zhou}}]{Abdikamalov:2020oci}%
  \BibitemOpen
  \bibfield  {author} {\bibinfo {author} {\bibfnamefont {A.~B.}\ \bibnamefont
  {Abdikamalov}}, \bibinfo {author} {\bibfnamefont {D.}~\bibnamefont
  {Ayzenberg}}, \bibinfo {author} {\bibfnamefont {C.}~\bibnamefont {Bambi}},
  \bibinfo {author} {\bibfnamefont {T.}~\bibnamefont {Dauser}}, \bibinfo
  {author} {\bibfnamefont {J.~A.}\ \bibnamefont {Garcia}}, \bibinfo {author}
  {\bibfnamefont {S.}~\bibnamefont {Nampalliwar}}, \bibinfo {author}
  {\bibfnamefont {A.}~\bibnamefont {Tripathi}}, \ and\ \bibinfo {author}
  {\bibfnamefont {M.}~\bibnamefont {Zhou}},\ }\href@noop {} {\  (\bibinfo
  {year} {2020})},\ \Eprint {http://arxiv.org/abs/2003.09663} {arXiv:2003.09663
  [astro-ph.HE]} \BibitemShut {NoStop}%
\bibitem [{\citenamefont {Blum}\ \emph {et~al.}(2009)\citenamefont {Blum},
  \citenamefont {Miller}, \citenamefont {Fabian}, \citenamefont {Miller},
  \citenamefont {Homan}, \citenamefont {van~der Klis}, \citenamefont
  {Cackett},\ and\ \citenamefont {Reis}}]{Blum:2009ez}%
  \BibitemOpen
  \bibfield  {author} {\bibinfo {author} {\bibfnamefont {J.}~\bibnamefont
  {Blum}}, \bibinfo {author} {\bibfnamefont {J.}~\bibnamefont {Miller}},
  \bibinfo {author} {\bibfnamefont {A.}~\bibnamefont {Fabian}}, \bibinfo
  {author} {\bibfnamefont {M.}~\bibnamefont {Miller}}, \bibinfo {author}
  {\bibfnamefont {J.}~\bibnamefont {Homan}}, \bibinfo {author} {\bibfnamefont
  {M.}~\bibnamefont {van~der Klis}}, \bibinfo {author} {\bibfnamefont
  {E.}~\bibnamefont {Cackett}}, \ and\ \bibinfo {author} {\bibfnamefont
  {R.}~\bibnamefont {Reis}},\ }\href {\doibase 10.1088/0004-637X/706/1/60}
  {\bibfield  {journal} {\bibinfo  {journal} {Astrophys. J.}\ }\textbf
  {\bibinfo {volume} {706}},\ \bibinfo {pages} {60} (\bibinfo {year} {2009})},\
  \Eprint {http://arxiv.org/abs/0909.5383} {arXiv:0909.5383 [astro-ph.HE]}
  \BibitemShut {NoStop}%
\bibitem [{\citenamefont {Wilms}\ \emph {et~al.}(2000)\citenamefont {Wilms},
  \citenamefont {Allen},\ and\ \citenamefont {McCray}}]{Wilms2000}%
  \BibitemOpen
  \bibfield  {author} {\bibinfo {author} {\bibfnamefont {J.}~\bibnamefont
  {Wilms}}, \bibinfo {author} {\bibfnamefont {A.}~\bibnamefont {Allen}}, \ and\
  \bibinfo {author} {\bibfnamefont {R.}~\bibnamefont {McCray}},\ }\href
  {\doibase 10.1086/317016} {\bibfield  {journal} {\bibinfo  {journal}
  {Astrophys. J.}\ }\textbf {\bibinfo {volume} {542}},\ \bibinfo {pages} {914}
  (\bibinfo {year} {2000})},\ \Eprint {http://arxiv.org/abs/astro-ph/0008425}
  {arXiv:astro-ph/0008425 [astro-ph]} \BibitemShut {NoStop}%
\bibitem [{\citenamefont {Steiner}\ \emph {et~al.}(2010)\citenamefont
  {Steiner}, \citenamefont {McClintock}, \citenamefont {Remillard},
  \citenamefont {Gou}, \citenamefont {Yamada},\ and\ \citenamefont
  {Narayan}}]{Steiner:2010kd}%
  \BibitemOpen
  \bibfield  {author} {\bibinfo {author} {\bibfnamefont {J.~F.}\ \bibnamefont
  {Steiner}}, \bibinfo {author} {\bibfnamefont {J.~E.}\ \bibnamefont
  {McClintock}}, \bibinfo {author} {\bibfnamefont {R.~A.}\ \bibnamefont
  {Remillard}}, \bibinfo {author} {\bibfnamefont {L.}~\bibnamefont {Gou}},
  \bibinfo {author} {\bibfnamefont {S.}~\bibnamefont {Yamada}}, \ and\ \bibinfo
  {author} {\bibfnamefont {R.}~\bibnamefont {Narayan}},\ }\href {\doibase
  10.1088/2041-8205/718/2/L117} {\bibfield  {journal} {\bibinfo  {journal}
  {Astrophys. J.}\ }\textbf {\bibinfo {volume} {718}},\ \bibinfo {pages} {L117}
  (\bibinfo {year} {2010})},\ \Eprint {http://arxiv.org/abs/1006.5729}
  {arXiv:1006.5729 [astro-ph.HE]} \BibitemShut {NoStop}%
\bibitem [{\citenamefont {Kulkarni}\ \emph {et~al.}(2011)\citenamefont
  {Kulkarni}, \citenamefont {Penna}, \citenamefont {Shcherbakov}, \citenamefont
  {Steiner}, \citenamefont {Narayan}, \citenamefont {Sadowski}, \citenamefont
  {Zhu}, \citenamefont {McClintock}, \citenamefont {Davis},\ and\ \citenamefont
  {McKinney}}]{Kulkarni:2011cy}%
  \BibitemOpen
  \bibfield  {author} {\bibinfo {author} {\bibfnamefont {A.~K.}\ \bibnamefont
  {Kulkarni}}, \bibinfo {author} {\bibfnamefont {R.~F.}\ \bibnamefont {Penna}},
  \bibinfo {author} {\bibfnamefont {R.~V.}\ \bibnamefont {Shcherbakov}},
  \bibinfo {author} {\bibfnamefont {J.~F.}\ \bibnamefont {Steiner}}, \bibinfo
  {author} {\bibfnamefont {R.}~\bibnamefont {Narayan}}, \bibinfo {author}
  {\bibfnamefont {A.}~\bibnamefont {Sadowski}}, \bibinfo {author}
  {\bibfnamefont {Y.}~\bibnamefont {Zhu}}, \bibinfo {author} {\bibfnamefont
  {J.~E.}\ \bibnamefont {McClintock}}, \bibinfo {author} {\bibfnamefont
  {S.~W.}\ \bibnamefont {Davis}}, \ and\ \bibinfo {author} {\bibfnamefont
  {J.~C.}\ \bibnamefont {McKinney}},\ }\href {\doibase
  10.1111/j.1365-2966.2011.18446.x} {\bibfield  {journal} {\bibinfo  {journal}
  {Mon. Not. Roy. Astron. Soc.}\ }\textbf {\bibinfo {volume} {414}},\ \bibinfo
  {pages} {1183} (\bibinfo {year} {2011})},\ \Eprint
  {http://arxiv.org/abs/1102.0010} {arXiv:1102.0010 [astro-ph.HE]} \BibitemShut
  {NoStop}%
\end{thebibliography}%
\end{document}